\documentclass{aa}  

\usepackage{txfonts}
\usepackage{graphicx,lipsum,overpic}
\usepackage{subfigure}
\usepackage{amssymb,amsmath}

\newcommand{\ignore}[1]{}

\def\aap {{A\&A}}

\def\apjl {{ApJL}}

\def\apjs {{ApJS}}

\begin{document}

   \title{Connecting planet formation and astrochemistry }

  \subtitle{A main sequence for C/O in hot-exoplanetary atmospheres}

   \author{Alex J. Cridland\inst{1}\thanks{cridland@strw.leidenuniv.nl}, Ewine F. van Dishoeck\inst{1,2}, Matthew Alessi\inst{3}, \& Ralph E. Pudritz\inst{3,4}
          }

   \institute{
   $^1$Leiden Observatory, Leiden University, 2300 RA Leiden, the Netherlands \\ $^2$ Max-Planck-Institut f\"ur Extraterrestrishe Physik, Gie{\ss}enbachstrasse 1, 85748 Garching, Germany \\ $^{3}$Department of Physics and Astronomy, McMaster University, Hamilton, Ontario, Canada, L8S 4E8 \\ $^4$Origins Institute, McMaster University, Hamilton, Ontario, Canada, L8S 4E8
             }

   \date{Received \today}


  \abstract
  {
To understand the role that planet formation history has on the observable atmospheric carbon-to-oxygen ratio (C/O) we have produced a population of astrochemically evolving protoplanetary disks. Based on the parameters used in a pre-computed population of growing planets their combination allows us to trace the molecular abundances of the gas that is being collected into planetary atmospheres. We include atmospheric pollution of incoming (icy) planetesimals as well as the effect of refractory carbon erosion noted to exist in our own solar system. We find that the carbon and oxygen content of Neptune-mass planets are determined primarily through solid accretion and result in more oxygen-rich (by roughly two orders of magnitude) atmospheres than Hot-Jupiters, whose C/O are primarily determined by gas accretion. Generally we find a `main-sequence' between the fraction of planetary mass accreted through solid accretion and the resulting atmospheric C/O - with planets of higher solid accretion fraction having lower C/O. Hot-Jupiters whose atmospheres have been chemically characterized agree well with our population of planets, and our results suggest that Hot Jupiter formation typically begins near the water ice line. Lower mass hot-Neptunes are observed to be much more carbon-rich (with 0.33 $\lesssim$ C/O $\lesssim$ 1) than is found in our models (C/O $\sim 10^{-2}$), and suggest that some form of chemical processing may affect their observed C/O over the few Gyrs between formation and observation. Our population reproduces the general mass-metallicity trend of the solar system and qualitatively reproduces the C/O-metallicity anti-correlation that has been inferred for the population of characterized exoplanetary atmospheres.
}

   \keywords{giant planet formation, astrochemistry
               }

    \maketitle
%

\section{Introduction}


The total bulk elemental abundances of the two most abundant atoms (after hydrogen and helium) have long been studied in the context of star and planet formation. Indeed cataloging the primary carriers of carbon and oxygen has been an important problem in astrochemical studies of the gas and dust in protoplanetary disks - the believed birth place of all planetary systems (see the review by \citealt{Pon14}).

The link between the astrochemistry in protoplanetary disks and planetary atmospheres was pointed out by \cite{Oberg11}. They argue that by measuring the carbon-to-oxygen ratio (C/O) in these atmospheres, one might work backwards to determine the location where the bulk of the planetary gas was accreted - since the gaseous (and solid) C/O varies with radii throughout the disk. In particular they note that the gas should have higher C/O than the Sun, while the ice C/O should be lower.

One might hope that the reverse might be true, that measuring the current C/O of a planetary atmosphere might tell us about the disk out of which the planet formed. However this belief is complicated by the theoretical result that planets migrate through their gas disks \citep{GoldTrem79,LP86,W91,MasPap03,Paard10,Paard11,Baru14}, they can be scattered by other planetary / stellar bodies once the disk has been evaporated \citep{Nice,Davies2014,Z15,Morbidelli2018}, and the carbon and oxygen budget can be transferred between the gas and ice by chemical reactions other than freeze-out and desorption \citep{Eistrup2016,Wei2019}. Hence a strong understanding of both the astrochemistry in the protoplanetary disk and planetary migration are necessary for interpreting the observed C/O in exoplanetary atmospheres. 

Currently, observational studies of atmospheric C/O have relied on retrieval techniques to infer the elemental abundances found in (predominantly) hot-Jupiters. These methods use C/O as an input parameter in forward chemical and structure models to compute a synthetic transmission / emission spectrum that is compared to the observed spectra (see for example the review by \citealt{Heng2018}).

Such obervational characterization studies of planetary atmospheres have inferred a wide range of possible C/O. Many of these studies have suggested solar-like C/O (=0.54) \citep{Moses13,Brogi2014,Line2014,Gandhi2018} or above \citep{Madu2012,Pinhas2019}. Some of these studies support carbon-rich atmospheres (C/O that exceed unity) \citep{Madu2012} which would be identified as having very water-poor atmospheres, and possibly having high abundance of HCN. Fewer works have suggested sub-solar C/O, but recently these ratios have begun to be found in smaller planets \citep{MacDonald2019}. Since the range of C/O between different stars can be quite wide, a good strategy is to compare the planetary C/O directly to the host star. In the cases where this has been done, the planetary C/O is found to be predominantly super-stellar \citep{Brewer16}. 

From a theoretical perspective, \cite{Crid17} showed that in the atmospheres of Hot Jupiters, C/O should tend towards the inherited C/O of the protoplanetary disk gas. This result was based exclusively on the atmospheric abundances being dominated by gas accretion alone, and the fact that each of their modelled planets migrated within their water ice line prior to accreting gas. Water has the highest sublimation temperature of all volatiles and hence within the water ice line the gas is `pristine', with little chemical evolution away from the initial volatile abundances in the disk. As outlined in \cite{Crid19b} this result implies sub-stellar C/O since the protoplanetary disk gas is expected to have sub-solar C/O, with the missing carbon being in refractory material. Alternatively, \cite{Mordasini16} found sub-stellar C/O in the atmospheres of their synthetic planets because of the accretion of mostly icy solid bodies. These icy bodies are always oxygen rich, because of the high H$_2$O and CO$_2$ abundances. 

\cite{Mordasini16} followed the work of \cite{Thiabaud2015} who similarly found that oxygen rich planetary atmospheres should result from the accretion of icy planetesimals. However both of these studies assumed that the gas was greatly depleted in carbon and oxygen due to the radial accretion of the gas through the disk. This assumption, however ignores the radial drift of icy pebbles and dust that would replenish the volatiles in the inner disk \citep{Booth2017,Bosman2017b,Booth2019}. Using simpler chemical prescriptions \cite{Mousis2011} argued that high C/O could not be directly accreted from the inner solar system, and \cite{AliDib2017} similarly found that a significant amount of core erosion was needed to explain high C/O.

Hence there is a general disconnect between atmospheric C/O inferred observationally to be super-stellar, and from the predictions of theory to be sub-stellar. An attempt was made in \cite{Crid19b} to lessen this discrepancy, by including an extra source of gaseous carbon which is generated by the chemical processing of carbon-rich dust grains \citep{Anderson2017,Gail2017,Klarmann2018}. In that work, it was found that this excess carbon did indeed improve the predictions of their theoretical model (with super-stellar C/O), if the chemical processes leading to the carbon excess was an ongoing process rather than a one-time release event.

In this work we extend the work of \cite{Crid19b} by combining this carbon excess model to the astrochemical results for a range of protoplanetary disk models and the planet formation within those disks. In doing so we will look to draw correlations between the underlying processes of planet formation and the resulting atmospheric C/O. Such a method is a common feature of planet population synthesis models and has been used in the past to learn more about planet formation than can be done by studying a single system.

Generally speaking, population synthesis incorporates the important physical processes of planet formation through a set of semi-analytic prescriptions. In this way, many planetary systems can be  constructed from a large range of initial conditions and physical parameters. With these synthetic populations of planets we can compare to the known population of planets in order to uncover the details of the underlying physics of planet formation and the chemistry in the disk (see for ex. \citealt{Benz2014}). As an example of this, \cite{AP18} investigate the impact of the envelope opacity on their synthetic populations of planets. They found that higher envelope opacities generally lead to an underproduction of `warm' Jupiters (Jupiter-mass planets at 1 AU) in their population. Hence they favour a lower envelope opacity for their future formation models (as is used here). Such a conclusion can only be made through the generation and comparison of synthetic populations of planets.

In what follows we take the formation histories for a sub-set of a synthetic population of planets and compute the chemical abundances of the gas and solids that are accreted in their atmospheres. In doing so we estimate the elemental abundances of carbon and oxygen in the atmospheres of these hot-Jupiters and super-Earths and compare to known exoplanets. We reproduce the known mass-metallicity relation \citep{MillerFortney2011,Kre14,Thorngren2016,Thorngren2018} observed in both the solar system and in known extra-solar planets, as well as derive a main-sequence of C/O, both dependent on the quantity of solid material that has accreted into the growing atmosphere.

We present our full model in sections \ref{sec:methods} \& \ref{sec:chem}. We compare our initial population of disk models to known population of protoplanetary disks (section \ref{sec:obsdisk}). We show our results of our combination of planet formation and astrochemistry in section \ref{sec:results}. These results are discussed and compared to observed atmospheric C/O in section \ref{sec:discussion} and conclude in \ref{sec:conclusion}.

\section{ Methods: Disk and planet formation model }\label{sec:methods}

For our purpose of estimating the bulk C/O of exoplanetary atmospheres, we require the combination of an evolving planet formation model with an evolving astrochemical disk model. Our planet formation model includes the planetesimal accretion paradigm and trapped planet migration and is featured in \cite{AP18}. The chemical model uses a modified version of the Michigan chemical code (as featured in \cite{Fogel11,Cleeves14,Crid16a,Schwarz2018} among others) and is described in section \ref{sec:chem}. Below we outline our semi-analytic disk model along with our planet formation model.

\subsection{ Protoplanetary disk model }\label{sec:diskmain}

Our gas and disk model rely on the semi-analytic model of \cite{Cham09} and the Two-population model of \cite{B12} respectively. Each of these have had small modifications in our previous work \citep{Crid16a,APC16a,Crid17} to account for processes like photoevaporation on the evolution of the gas disk and the impact of a deadzone on the settling and growth of the dust grains. The disk is described by a 1+1D model, where the temperature and density of the gas is described by a power-law of radius and mass accretion rate ($\dot{M}$) and $\dot{M}$ depends on time. We outline the details of these models in Appendix \ref{sec:disk}.

\subsection{ Planet formation }\label{sec:form}

Planet formation is a complicated process - with many underlying assumptions and parameters. To better understand it, large populations of synthetic planets are compared to the known population in order to learn which aspects of planet formation best determine a planet's final properties. 

As seen in \cite{Mordasini2009a,Mordasini2009b}, \cite{HP13}, \cite{Chambers2018}, and \cite{AP18} the common observables used to constrain the population are often planetary mass and orbital radius (assuming circular orbits). In this work we wish to expand to a third dimension of comparison - the bulk chemical composition of the planetary atmosphere. To build these synthetic populations of planets we randomly select different initial conditions and/or model parameters from underlying distributions (more in \ref{sec:init}) and compute the resulting planetary growth.

Our planetary growth model relies on the planetesimal accretion paradigm \citep{KI02,IL04a} which assumes that the initial planetary embryos are built through subsequent accretion of 10-100 km planetesimals. Specifically we use a subset of a population of synthetic planets from Alessi, Pudritz, \& Cridland (APC, in prep). This subset includes hot Jupiters, defined by orbital radii $< 0.1$ AU and planetary mass larger than 10 M$_\oplus$. Additionally the sub-population includes close in super-Earths with a similar orbital distribution as the hot Jupiters, but with $M_p<$ 10 M$_\oplus$. As they grow the proto-planets migrate through the disk. We assume that their migration is described by the `planet trap' paradigm which assumes that migration is limited by discontinuities in disk properties. The details for both of these processes can be found in Appendix \ref{sec:plntform}.

\ignore{
\subsubsection{ Building a population: Log-normal distribution of initial conditions }\label{sec:init}
}

\subsection{ Population synthesis: building a population from a log-normal distribution of initial conditions }\label{sec:init}

Presumably the initial conditions leading to the formation of planets vary from system to system. While the underlying distribution or physical constraints are not fully understood, current observational efforts have worked to provide limits. In their population synthesis model \cite{AP18} assumed that their initial conditions and parameters could be drawn from log-normal distributions (for $M_{disk,0}$ and $t_{life}$) and a normal distribution (for $[Fe/H]$) for which they prescribed an average value and standard deviation. The exception to this was the selection of $f_{max}$ which was drawn from a distribution of equal probability.

The three disk parameters that were selected from log-normal distribution were the initial disk mass ($M_{disk,0}$), the disk lifetime ($t_{life}$), and the disk metallicity ($[Fe/H]$). These parameters impacted the initial gas surface density distribution through eq. \ref{eq:15}, the depletion time ($t_{dep}$ in eq. \ref{eq:19}), and the initial (global) dust-to-gas ratio ($f_{dtg}$) used in the Two-population model respectively. The dust-to-gas ratio was used to initialize the dust simulation such that $\Sigma_d(r,t=0) = f_{dtg}\Sigma_g(r,t=0)$. The dust-to-gas ratio was set following \citep{AP18}:\begin{align}
f_{dtg} = f_{dtg,0}10^{[Fe/H]},
\end{align}
where $f_{dtg,0}= 0.01$ is the Interstellar dust-to-gas ratio that we assume is representative of solar abundances: $[Fe/H]_{\rm solar} \equiv 0$. Over the lifetime of the disk, the dust-to-gas ratio evolves due to the radial drift of dust (see Appendix \ref{sec:dust}).

\begin{table}
\caption{Parameters for the distributions which are used to select the initial conditions in the APC population. }
\centering
\begin{tabular}{| l | c | c |}
\hline
 & Average & 1$\sigma$ range\\\hline
$M_{disk,0}$ ($M_\odot$) & 0.1 & 0.073 -- 0.137 \\\hline
$t_{life}$ (Myr)  & 3 & 1.8 -- 5 \\\hline
  $[Fe/H]$   & -0.02 & -0.22 -- 0.18 \\\hline
\end{tabular}
\label{tab:01}
\end{table}

The population we use in this work is a subset of the population from APC, where the initial disk mass, disk lifetime, and metallicity are varied. The average values and 1$\sigma$ range of these parameters are listed in table \ref{tab:01}.

\begin{figure}
\centering
\includegraphics[width=0.5\textwidth]{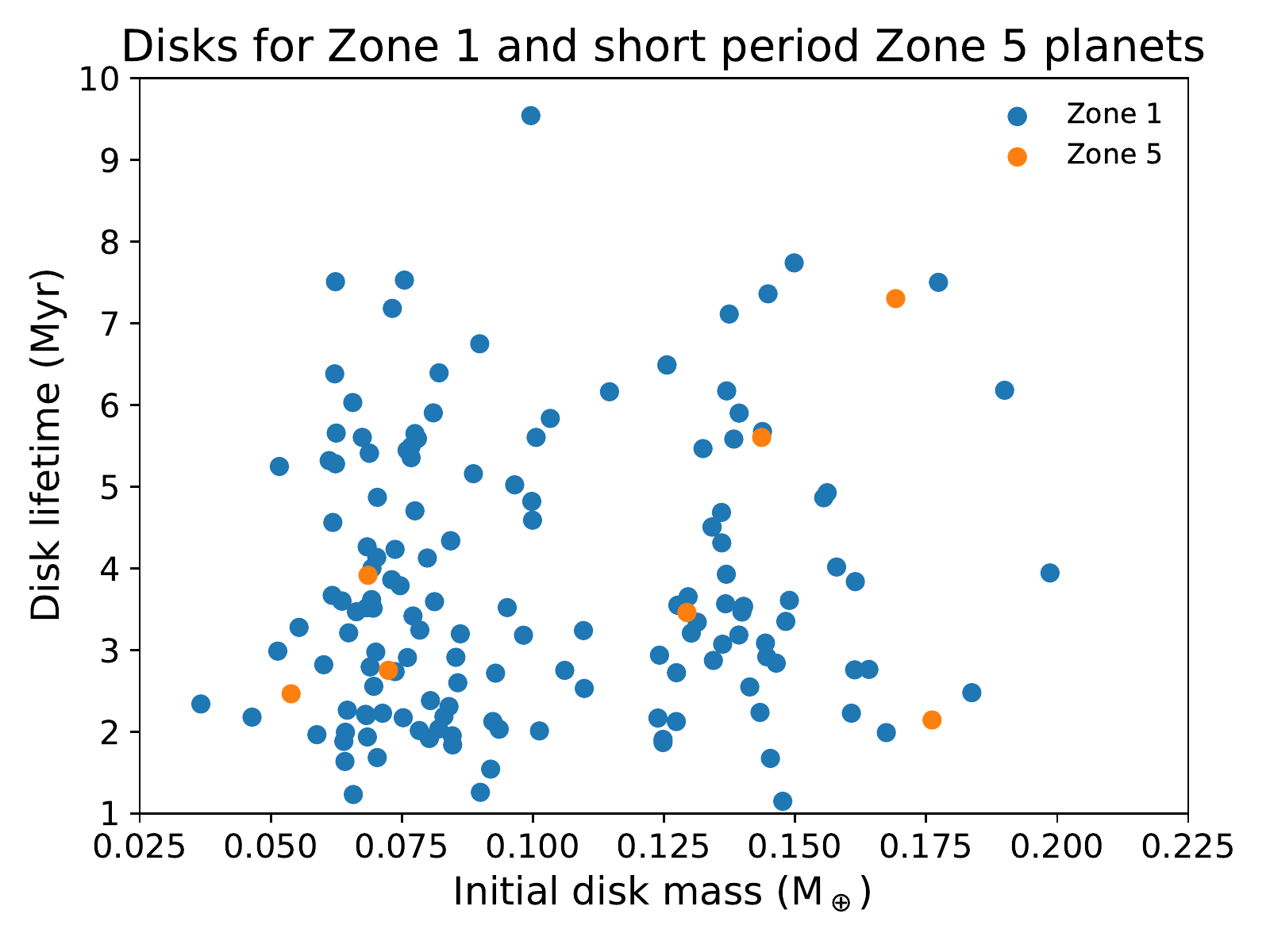}
\caption{Range of disk initial masses and lifetimes for the planet sub-population from APC used in this work.}
\label{fig:meth01}
\end{figure}

In figure \ref{fig:meth01} we show the distribution of disk initial mass and lifetime that generates the sub-population of planets used in this work. We note that our population is dominated (i.e. 151 of 158) by hot-Jupiters (Zone 1 planets from \cite{AP18}) as well as 7 less massive Zone 5 planets. We find that planets coming from very long lived disks (age $>$ 8 Myr) are rare, simply because those long lived disks are rarer in our distribution. As are planets from the very low mass ($< 0.025$ M$_\odot$) and very high mass ($> 0.2$ M$_\odot$) disks.

\subsection{Comparing disk mass distribution to known systems}\label{sec:obsdisk}

\begin{figure*}
\centering
\includegraphics[width=\textwidth]{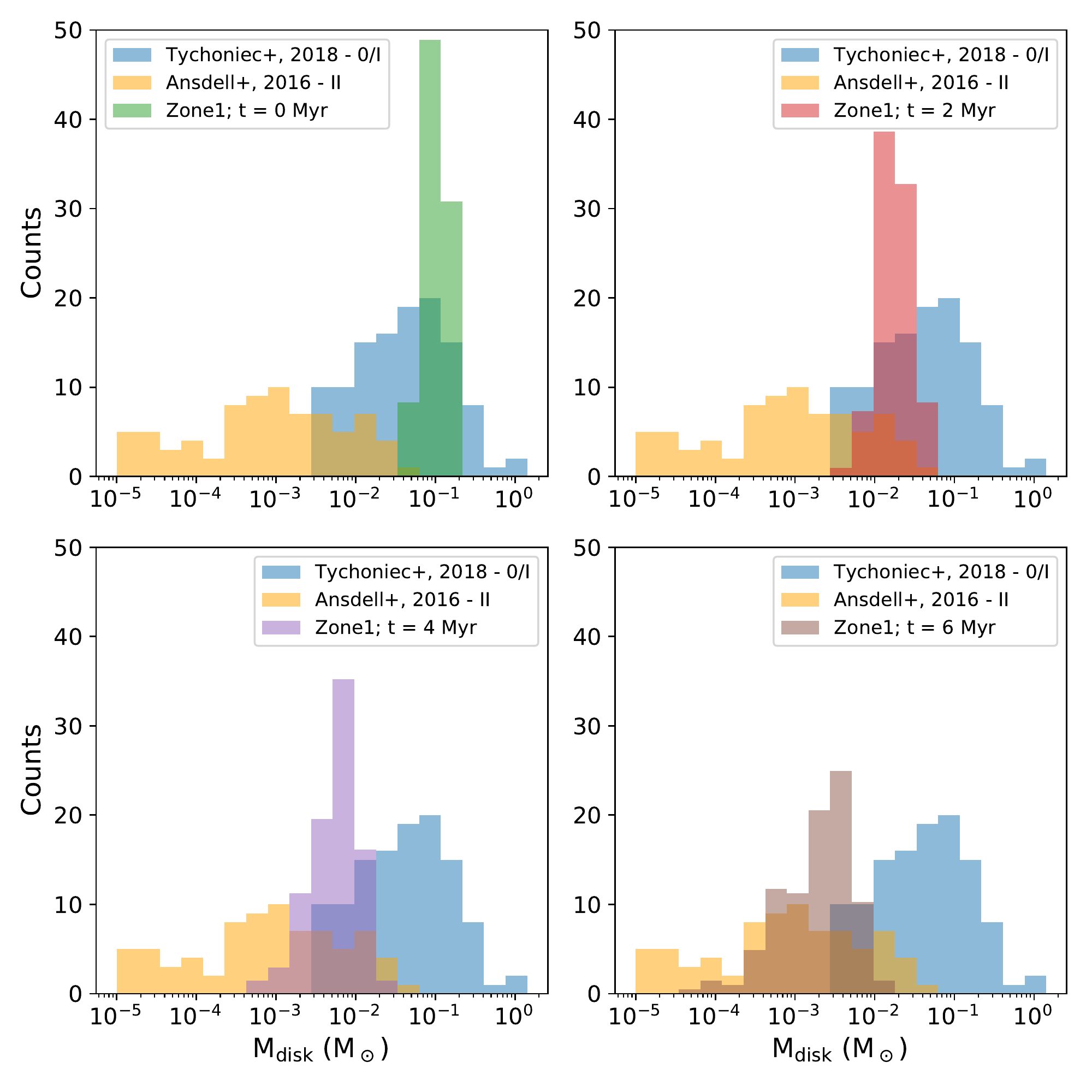}
\caption{A comparison of the distribution of disk masses in our population of Zone 1 planets, evolved using the model presented in section \ref{sec:gas}. We show the mass of each of the disk models at its initial time, 2, 4, and 6 Myr into their evolution, even if the disk lifetime for any particular disk model is shorter than those times. We compare to the disk masses measured in a population of class 0/I from \citet{Tychoniec2018}, as well as a population of class II objects from \citet{Ansdell2016}.  }
\label{fig:meth02}
\end{figure*}

In figure \ref{fig:meth02} we compare the mass range of our protoplanetary disk model at 4 times through its evolution (0, 2, 4, 6 Myr) against the inferred masses of young (class 0/I) and intermediate (class II) aged systems. The young systems are from \cite{Tychoniec2018}, while the intermediate aged systems are taken from \cite{Ansdell2016}. For the class II systems we took the published dust mass and multiplied by the standard ISM value of the gas-to-dust ratio (100). This is in line with the assumption of \cite{Tychoniec2018} and ignores the possible discrepancy between observed CO fluxes and the total disk mass \citep{Farve2013,Ansdell2016,Bosman2018,Krijt2018,Schwarz2018,Schawrz2019,BoothA2019}.

In comparing to observations we see that the initial mass range (green) that we use is much less broad than is seen in the young population of systems seen by \cite{Tychoniec2018} (light blue) . This includes primarily class 0/I objects with less gas mass than is included in our population. While we are naturally biased towards forming giant planets this discrepancy does suggest that the lower mass systems seen in the observations may not lead to the formation of hot-Jupiters and close in Zone 5 planets. 

As our models evolve through 2-4 Myr we find that our distribution overlaps with the higher mass end of the \cite{Ansdell2016} population of class II disks. Since planet formation via planetesimal accretion is generally slow (see for ex. \citealt{Bitsch2015}) we require larger mass disks to build Jupiter-mass objects within the lifetime of the disks. Hence based on the planetesimal paradigm, these lower mass disks will not generate large planets within the lifetime of the disk.

At the latest (6 Myr) stages our mass distribution begins to spread due to differences in the photoevaporation timescale $t_{dep}$ in eq. \ref{eq:19}. We still remain mostly at the large end of the observed mass distribution. This result is likely due to our aforementioned bias towards the most massive disks.

\section{ Methods: Disk astrochemistry }\label{sec:chem}

\subsection{ Volatile chemistry }

As the temperature and density structure of the disk evolves over its lifetime, so do the molecular abundances of volatiles in the gas and ice. This is different from the refractory component of the disk, locked up in dust grains that remains (mostly) chemically neutral (but see below). As in our previous work \citep{Crid16a,Crid17} we use the Michigan chemical code \citep{Fogel11,Bethell11,Cleeves14} to compute the chemical kinetics of the gas and ice. 

The Michigan chemical code computes the chemical kinetics of the gas and icy dust grains. The chemical network is predominantly in the gas-phase, driven mostly by ions produced by cosmic ray ionization and interactions with the high energy radiation fields described in section \ref{sec:field}. Along with the ion and neutral gas-phase reactions there are a set of dust grain surface reactions that involve the freeze out and desorption of volatiles and the production of H$_2$O and H$_2$ on the dust grain. The chemical network is based on the OSU gas-phase network \citep{Sea04} and includes additions made by \cite{Fogel11} to include photodissociation, CO and H$_2$ self-shielding, and non-thermal cosmic-ray ionization of H$_2$ and helium.

As discussed above, we compute the growth and radial drift of the dust grains.  The primary impact of the dust grain evolution changes the average size of the dust grains available for volatiles freeze-out. As the dust grains grow and radially drift in, the average surface area of the dust grains is reduced, which slows the freeze-out of volatiles in the outer disk. However unlike \cite{Booth2019} we do not couple the chemistry and dust evolution to account for the radial drift of icy volatiles. While this dynamical evolution was shown to may have an important impact on the local C/O of the gas and dust (\citealt{Booth2019}, but also see \cite{Krijt2016,Booth2017,Bosman2017b}), it remains to be seen what impact this evolution has when coupled to the dynamic nature of planet formation (i.e. planetary migration also radially evolves the growing planet). We thus leave the implication of radially drifting dust on the local gas C/O to future work, and instead focus on the impact that smoothly varying the average size of the dust grains have on chemical processes like volatile freeze out and desorption (see \citealt{Crid17}).

Previously, we accounted for any change in the molecular abundance of the gas in the disk by selecting many hundred snapshots of the disk's gas temperature and density evolution, then computed the chemical kinetics for 1 Myr on each of the snapshots \citep{Crid17}. The molecular abundance was initialized at the same abundance for each snapshot, and the chemical kinetics are computed for long enough to insure that the chemical system had run to a steady state. As a result the source of any chemical change is attributed to differences in the gas / dust properties between different snapshots.

Running the Michigan code in this `passive disk' method meant that running the chemical evolution over the lifetime of the disk was slow and spending roughly 2-3 months on a single astrochemical disk model meant that only a few disk models could be run at a time. Hence building a large population of models to test planet formation was not possible in any meaningful time frame.

A second side effect of the `passive disk' method was that by beginning each snapshot with the same initial state, we erased any evolution that had built up in previous snapshots. While we did confirm that differences that arose from initializing a snapshot with the results of the previous snapshot were small, they could build up as hundreds of snapshots are run. Due to the complexity of these large chemical systems a more natural approach is to allow the chemistry to evolve simultaneously with the gas and dust properties (temperature and density).

\begin{figure}
\centering
\includegraphics[width=0.5\textwidth]{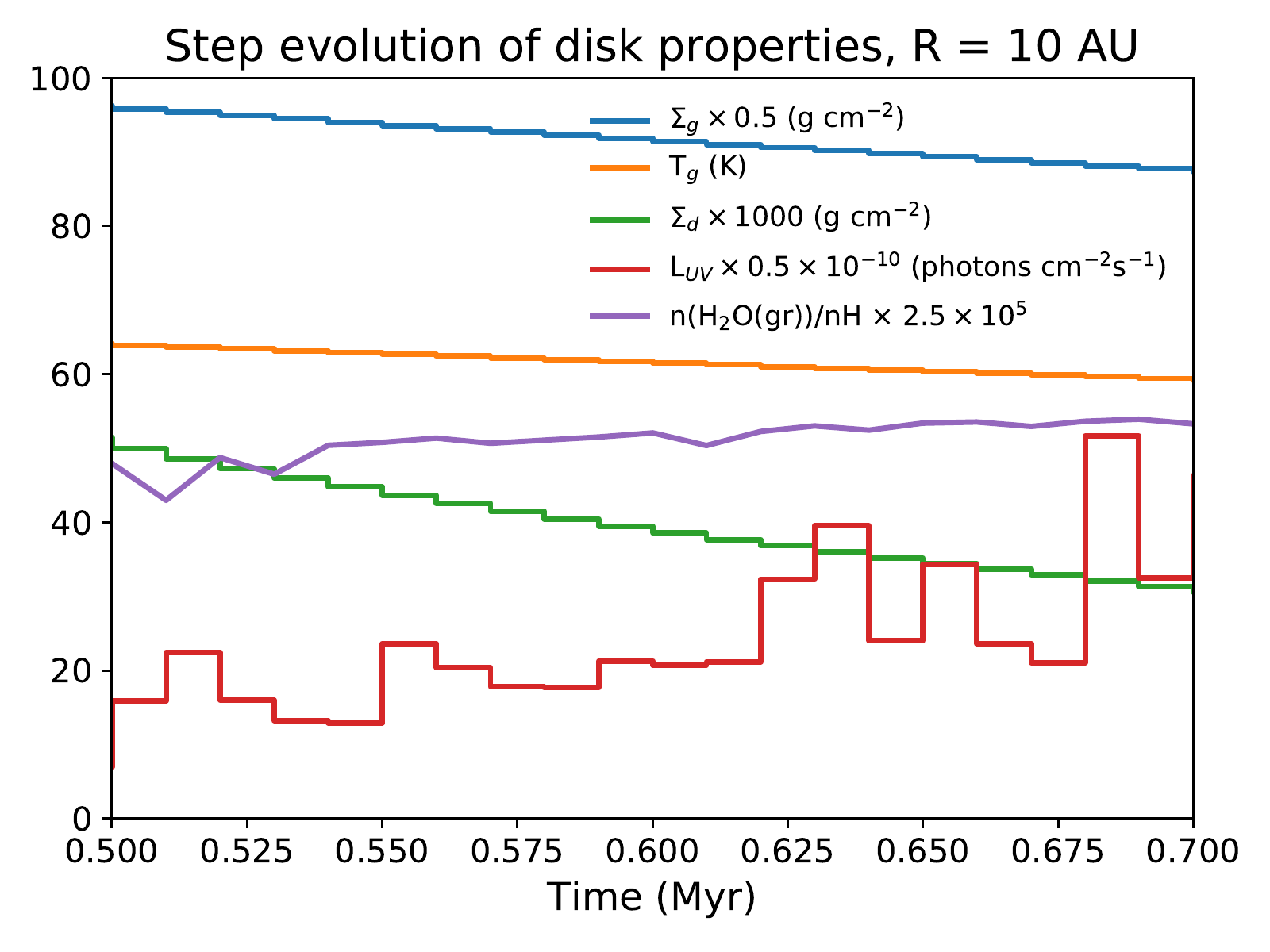}
\caption{An example of how the physical properties of the disk evolve between 0.5 and 0.7 Myr at a radius of 10 AU. The gas temperature, abundance of water ice and UV flux are measured at the disk midplane. The gas temperature and density, the dust density, and the flux high energy radiation stay constant during a single time step. They are then adjusted to a new value at the end of the timestep before the chemical evolution moves onward in time.}
\label{fig:meth03}
\end{figure}

For this and subsequent work, we have improved on the original Michigan code by allowing the underlying disk model (gas / dust temperature and density) to evolve along with the chemistry. To be clear - there is no back reaction between the chemistry and gas properties. The disk evolution is pre-computed using the methods described in section \ref{sec:disk}, then are fed into the chemical code. We generate snapshots of the gas temperature and density which cover the entire evolution of the disk. These snapshots are temporally spaced by $10^4$ years, which sets the length of time over which the disk properties remain constant (see figure \ref{fig:meth03}). Keeping the disk properties constant in this way supports the other disk properties (dust distribution, high energy radiation field) that are also pre-computed, and are too computationally expensive to change continually.

In figure \ref{fig:meth03} we show an example of how the temperature and density structures evolve with time. Between time steps the physical properties of the disk remain constant, while the chemistry evolves. Then after the time step elapses the physical properties are changed and the chemistry continues. For comparison we note the evolution of water ice at 10 AU, which shows only minimal changes over the short time frame shown here.

\begin{table}
\centering
\caption{Initial abundances relative to the number of H atoms. Included is the initial ratio of carbon to oxygen (C/O) and the initial ratio of carbon to nitrogen (C/N). These are the same as used in \cite{Crid17} and are based on \cite{Fogel11}. We similarly show the elemental abundance of the carbon and oxygen in the refractories (C$_{\rm ref}$ and O$_{\rm ref}$). }\label{tab:initchem}
\begin{tabular}{l l l l}\hline\hline
Species & Abundance & Species & Abundance \\\hline
H$_2$ & $0.5$ & H$_2$O & $1.5\times 10^{-4}$ \\
He & $0.14$ & N & $2.25\times 10^{-5}$ \\
CN & $6.0\times 10^{-8}$ & H$_3^+$ & $1.0\times 10^{-8}$ \\
CS & $4.0\times 10^{-9}$ & SO & $5.0\times 10^{-9}$ \\
Si$^+$ & $1.0\times 10^{-11}$ & S$^+$ & $1.0\times 10^{-11}$ \\
Mg$^+$ & $1.0\times 10^{-11}$ & Fe$^+$ & $1.0\times 10^{-11}$ \\
C$^+$ & $1.0\times 10^{-9}$ & HCO$^+$ & $9.0\times 10^{-9}$ \\
CO & $1.0\times 10^{-4}$ & N$_2$ & $1.0\times 10^{-6}$ \\
C & $7.0\times 10^{-7}$ & NH$_3$ & $8.0\times 10^{-8}$ \\
HCN & $2.0\times 10^{-8}$ & C/O & $0.40$ \\
C$_2$H & $ 8.0\times 10^{-9}$ & C/N & $4.09$ \\
C$_{\rm ref}$ & 2.44$\times 10^{-4}$ & O$_{\rm ref}$ & 1.75$\times 10^{-4}$\\
\hline
\end{tabular}
\end{table}

In table \ref{tab:initchem} we show the initial molecular abundances used in the astrochemical simulations, we additionally note the C/O and C/N in the gas disk. The individual elemental ratios (excluding refractories) are C/H $\sim 1.0\times 10^{-4}$, O/H $\sim 2.5\times 10^{-4}$, and N/H $\sim 2.45\times 10^{-5}$.

\begin{figure}
\centering
\includegraphics[width=0.5\textwidth]{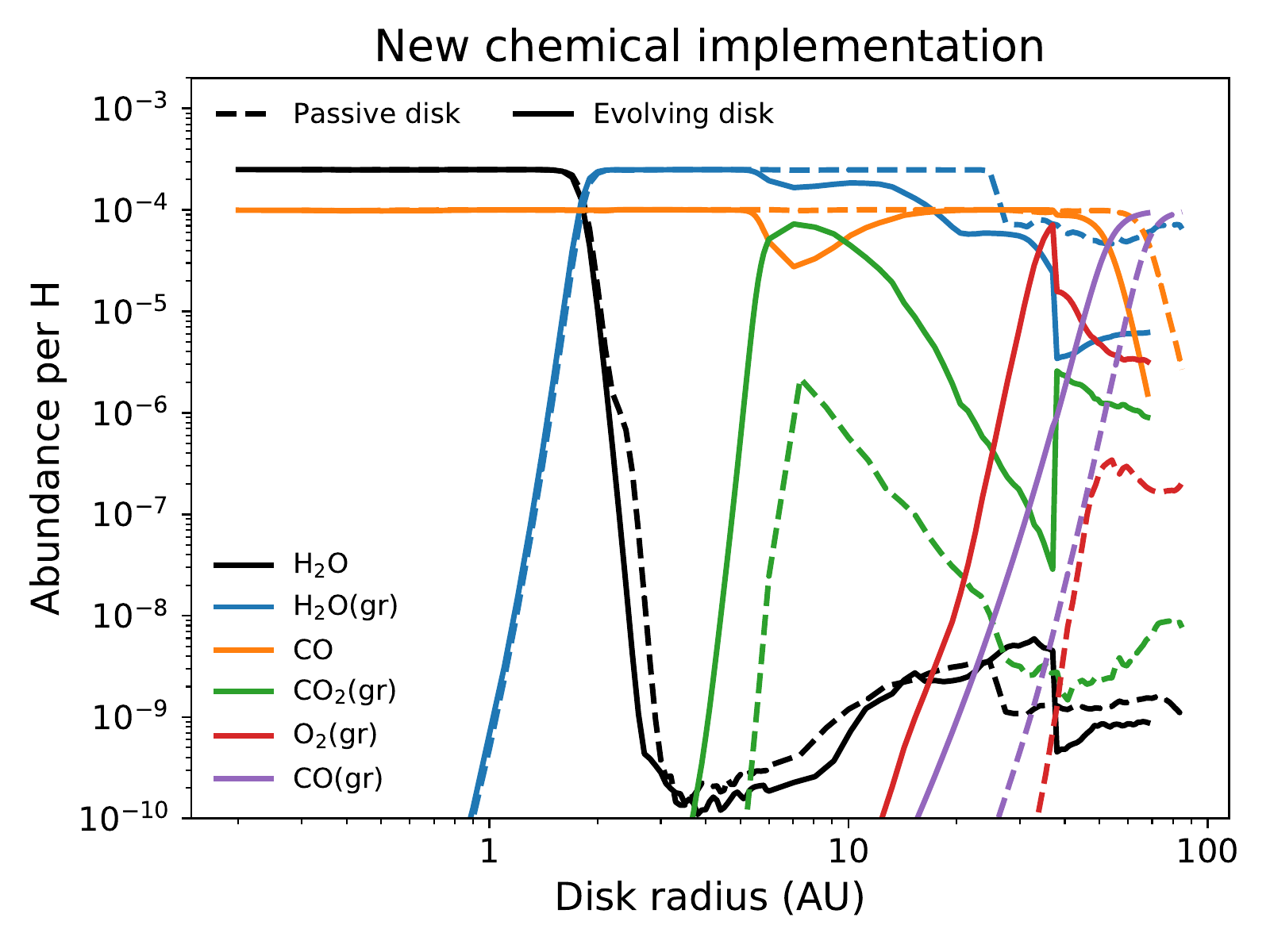}
\caption{Midplane distribution of the primary carbon and oxygen bearing volatile molecules as computed by the original Michigan chemical code (Passive disk) and our modified version (Evolving disk). The label `(gr)' denotes molecules that are frozen on the dust grains. Each model is run up to 1 Myr, however in the Evolving model the disk temperature and gas surface density evolves with time.}
\label{fig:newchem}
\end{figure}

\begin{figure*}
\centering
\includegraphics[width=\textwidth]{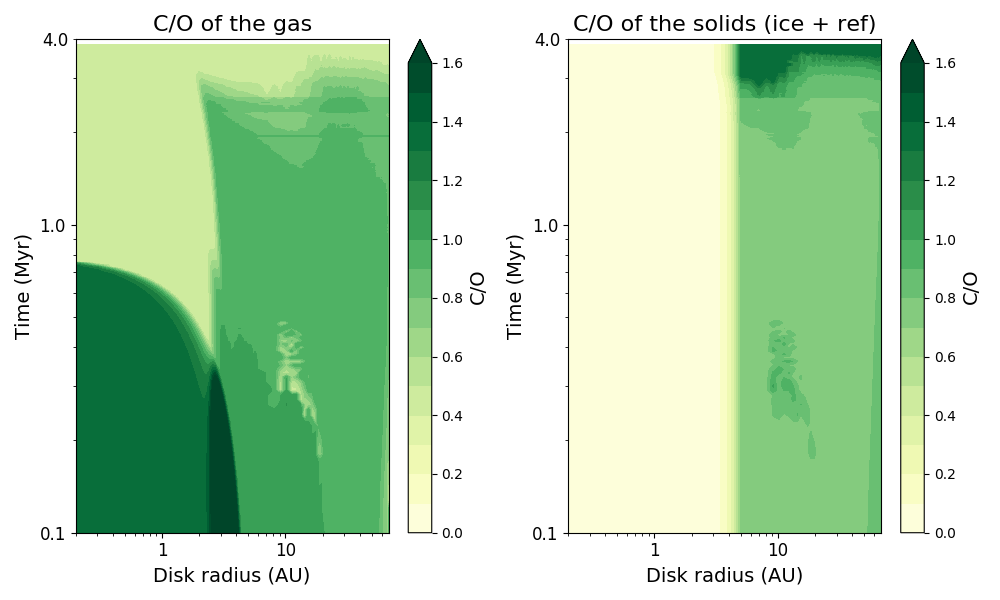}
\caption{ Carbon to oxygen ratio (C/O) of the gas and dust (ice and refractories) for a disk model in our population. The gas disk has a higher C/O than unity for $\sim 0.5$ Myr until the disk has cooled slightly. The ice is almost exclusively oxygen rich, apart from a region of the inner disk at late times caused by the slow production and freeze out of long chain hydrocarbons. Here we have included the excess carbon from the `reset' model, which eventually disappears after nearly 0.8 Myr. The solids in the region of the disk where the carbonaceous dust is eroded is always oxygen rich, since the silicates remain in the dust grain while the carbon is removed. C/O changes most at the ice lines of the disk, particularly the water ice line between 2 and 3 AU. The CO ice line is located outside of 60 AU, and hence does not show up on the figure.}
\label{fig:ctoomap}
\end{figure*}

In figure \ref{fig:newchem} we compare the radial distribution of molecular abundances (along the midplane of the disk) for a few abundant species, which are computed by both the Passive disk method (dashed lines) and the Evolving disk method (solid line) at a time of 1 Myr. We find only minimal difference in the chemical abundances in the inner (r $<$ 4 AU) disk, while farther out we do begin to see a difference in the computed abundances. In particular we see a build up of frozen CO$_2$ and O$_2$ (denoted by `(gr)' in the figure) in the outer disk. 

 The production of these molecules occurs (in our model) in the gas phase through a reaction of CO and the OH radical (for CO$_2$) and elemental O with OH (for O$_2$). In both cases the OH radical is required, which can be produced by the dissociation of H$_2$O, which explains the reduction of H$_2$O abundance in the same part of the disk where CO$_2$ and O$_2$ become more abundant. This dissociation is driven by cosmic-ray induced UV photons and is quite slow (assuming the typical cosmic-ray ionization rate of $\sim 10^{-17}~s^{-1}$), requiring timescales on the order of a Myr to reach the reduction we see in the figure (see \citealt{Eistrup2016} for t $>$ 1 Myr of evolution). Hence by continually resetting the water abundances in each snapshot (as was done in the Passive disk method) we were restricting these reaction pathways. Regardless, this resetting has little impact on our previous results, since every modelled planet accreted gas either near or within the water ice line within 2 AU at 1 Myr.

We note that in our model the carbon budget in the outer regions of the disk is dominated by CO. This is due to the fact that in our chemical model we do not include dust grain surface reactions like the hydrogenation of CO ice to produce methanol \citep{Drozd14,Walsh2014,Eistrup2018}. Since planet formation occurs far inward of the CO ice line, omitting these reactions should not greatly impact the chemical compositions of our forming planets.

In Figure \ref{fig:ctoomap} we show C/O in one of the disk models that we use in our population. The evolution of C/O here is simpler than was shown in \cite{Crid19a} because we have ignored, for simplicity, the majority of gas-grain chemical reactions which are included in \cite{Eistrup2018}. Similarly we see less evolution in the C/O of the ice species, which stays low over nearly all radii and time. An exception of this is in the inner disk at late times where the production and freeze out of hydrocarbons eventually occurs, producing a low abundance of carbon species on the icy grains. Since the oxygen abundance of the ice is already small at these radii, the low abundance of these frozen carbon species result in these high C/O.

\subsection{ Refractory carbon erosion }\label{sec:refc}

As was alluded to earlier, the majority of the refractory component of the disk - the dust and larger solid bodies - does not undergo significant chemical change over the lifetime of the disk. An exception occurs in the inner few AU and has come to be known as refractory carbon erosion \citep[called `depletion' in these works: ][]{Berg15,Anderson2017,Gail2017,Klarmann2018,Crid19b}. To avoid confusion we will denote this processes `erosion' since `depletion' is an astrochemical term generally connected to a reduction in gas phase volatile abundances due to its incorporation into the dust grains - where here we will describe the opposite effect.

In the solar system the Earth shows a reduction in the bulk carbon relative to silicon of about three orders of magnitude relative to the Interstellar medium (ISM), while belt asteroids show a range of erosion between one and two orders of magnitude \citep{Berg15}. This observation encouraged \cite{Mordasini16} to suggest a universality to this refractory carbon erosion and include it in their planet formation models.

More recently, \cite{Crid19b} included the chemical impact of carbon refractory erosion on the accretion of carbon in their formation model, but also included the excess {\it gaseous} carbon that would be released by the physical or chemical process responsible for the eroded carbon. A main result of their work was to show that this extra source of gaseous carbon in the inner disk could drive the carbon-to-oxygen ratio (C/O) of Hot Jupiter atmospheres as high as $\sim$ 2.5 times their value when the carbon excess is ignored.

Here we follow the same methods as in \cite{Crid19b}. First they derive a functional form to describe the refractory carbon-to-silicon ratio (C/Si) in the disk based on figure 2 of \cite{Mordasini16}: \begin{align}
{\rm C/Si}_{\rm ref}(r) = \begin{cases}
0.001 & r_{\rm AU} \le 1 \\
0.001 r_{\rm AU}^c & 1 < r_{\rm AU} < 5 \\
6 & r_{\rm AU} \ge 5, \\
\end{cases}
\label{eq:refc01}
\end{align}
where $c \sim 5.2$ is needed to connect the inner ($<1$ AU) disk to the outer ($>5$ AU) disk and $r_{AU}$ is the orbital radius in units of AU.

In the ISM, \cite{MishraLi2015} report Si/H$_{\rm ISM}$ $= 40.7$ ppm which when combined with C/Si$_{\rm ISM} = 6$ \citep{Berg15} results in a refractory carbon abundances of C/H$_{\rm ISM} = 2.44\times 10^{-4}$. With that in mind the excess gaseous carbon in the disk would be: \begin{align}
{\rm C/H}_{\rm exc}(r) &= \left({\rm C/Si}_{\rm ISM} - {\rm C/Si}_{\rm ref}(r)\right){\rm Si/H}_{\rm ISM} \nonumber\\
{\rm C/H}_{\rm exc}(r) &= \begin{cases}
2.44\times 10^{-4} & r_{\rm AU} \le 1 \\
2.44\times 10^{-4}\left( 1 - 0.000167 r_{\rm AU}^{c}\right) & 1 < r_{\rm AU} < 5 \\
0 & r_{\rm AU} \ge 5. \\
\end{cases}
\label{eq:refc02}
\end{align}

A final level of complexity comes from the source of the refractory carbon erosion. This source is still not well understood and is complicated by the vertical mixing of the dust grains \citep{Anderson2017} as well as their radial drift \citep{Klarmann2018}. Here we make the same simplification as in \cite{Crid19b} by requiring that the source of the excess carbon was {\it either} an ongoing process like oxidation or photodissociation (as in \citealt{Anderson2017}) or was the result of a thermal event early in the disk lifetime that released the carbon from the grains \citep{Berg15,Gail2017}. We differentiate these two model as `ongoing' and `reset' respectively.

In the case of the `reset' model, the excess carbon would be released very early in the disk's life and hence would advect towards the host star with the bulk of the gas. In the case of the `ongoing' model the same advection would occur, but since the dust radially drifts into the erosion region and is processed, the excess carbon is replenished. Functionally the `reset' model has the form:\begin{align}
&{\rm C/H}_{\rm exc}(r,t) =\nonumber\\
&\begin{cases}
2.44\times 10^{-4} & r_{\rm AU} \le 1 - v_{\rm adv}t \\
2.44\times 10^{-4}\left( 1 - 0.000167 r_{\rm AU}^{c}\right) & 1 - v_{\rm adv}t < r_{\rm AU} < 5 - v_{\rm adv}t \\
0 & r_{\rm AU} \ge 5- v_{\rm adv}t, \\
\end{cases}
\label{eq:refc03}
\end{align}
where $v_{adv} = 6.32$ AU Myr$^{-1}$ is the net advection speed of the bulk gas taken from \cite{Bosman2017b} for a disk with $\alpha = 0.001$. In the case of the `ongoing' model we set $v_{adv} = 0$.

In our chemical models we do not include the excess carbon during the chemical kinetics simulation. Instead we simply accrete C/H$_{\rm exc}$ into the planetary atmosphere on top of the elemental carbon that is naturally available from the disk gas (see table \ref{tab:initchem}). Our assumption is that the excess carbon would remain primarily in the gas phase, producing CO (at the expense of H$_2$O), HCN, or CH$_4$. However within 5 AU, \cite{Wei2019} show that long chain hydrocarbons can reform on the ice mantle with an abundance of approximately 1\% of the total carbon elemental abundances. While we neglect the chemical impact of the carbon-excess here, we will account for the effect on atmospheric C/O by grain surface chemical reactions in future work.

The primary refractory source of oxygen is stored in silicates with SiO$_3$ and SiO$_4$ functional groups. To compute the elemental abundance of oxygen we follow \cite{Mordasini16}, who assume a refractory mass ratio of 2:4:3 for carbon, silicates, and iron respectively. This assumption leads to a refractory abundance O/H$_{\rm ref} = 1.75\times 10^{-4}$.

Before continuing, we note that the above physical processes describing the erosion of refractory carbon is separate from the ongoing discussion regarding depleted gaseous CO in the cold outer disk (r $>$ 10 AU) suggested in recent ALMA surveys \citep{Favre2013,Kama2016,Miotello2016,Yu2017,Bergin2018,Cleeves2018,Schwarz2018}. The process that we have described here is concentrated in the inner 5 AU of the disk which will not have an effect on the observable volatile carbon abundance in the gas of the outer disk. We have however, ignored the impact of radial delivery of carbon and oxygen carring volatiles by drifting pebbles as reported by \cite{Booth2017}, \cite{Bosman2017b}, and \cite{Booth2019}. Hence the physical processes that are depleting the volatile carbon abundance in the outer parts of the disk could increase the carbon abundance in the inner solar system. This connection is computationally expensive to include in our model, and hence we leave its impact to further work.

\subsection{ Accounting for the accretion of ices and refractories }\label{eq:chemacc}

As is discussed in \cite{Crid19b} we assume that no refractory carbon or oxygen that accretes into the core is incorporated into the atmosphere, by either erosion or out-gassing. We assume that once the gas envelope is sufficiently massive (more than 3 M$_\oplus$, see below), planetesimals can begin to deliver ices and refractories. 

\cite{Mordasini15} compute the survivability of planetesimals as they pass through a planetary atmosphere. This calculation is complicated and involves self-consistent models of planet formation, atmospheric structure, and thermal processing of the incoming planetesimals. 

We strive to capture the essence of these more complicated models by assuming that below an atmospheric mass of 3 M$_\oplus$ the refractory component of the planetesimal remains intact through the atmosphere and is incorporated into the core. This cut off is based on the calculation done by \cite{Mordasini15}, and represents a minimum atmospheric mass above which planetesimals of all sizes are evaporated within the atmosphere. During their trip through the atmosphere, these planetesimals are heated as they pass through the gas releasing any frozen volatiles into the atmosphere. We assume that a planetesimal releases its entire volatile component upon entering an exoplanetary atmosphere (of any mass).

When the mass of the atmosphere grows above a mass of 3 M$_\oplus$, planetesimals are completely evaporated in the atmosphere; releasing their volatiles and refractories into the gas. We assume efficient mixing such that these planetesimals incorporate their refractory carbon and oxygen into the bulk C/O.

\section{ Results: Chemical population synthesis }\label{sec:results}

\subsection{ Population of planets }\label{sec:population}

\begin{figure}
\centering
\includegraphics[width=0.5\textwidth]{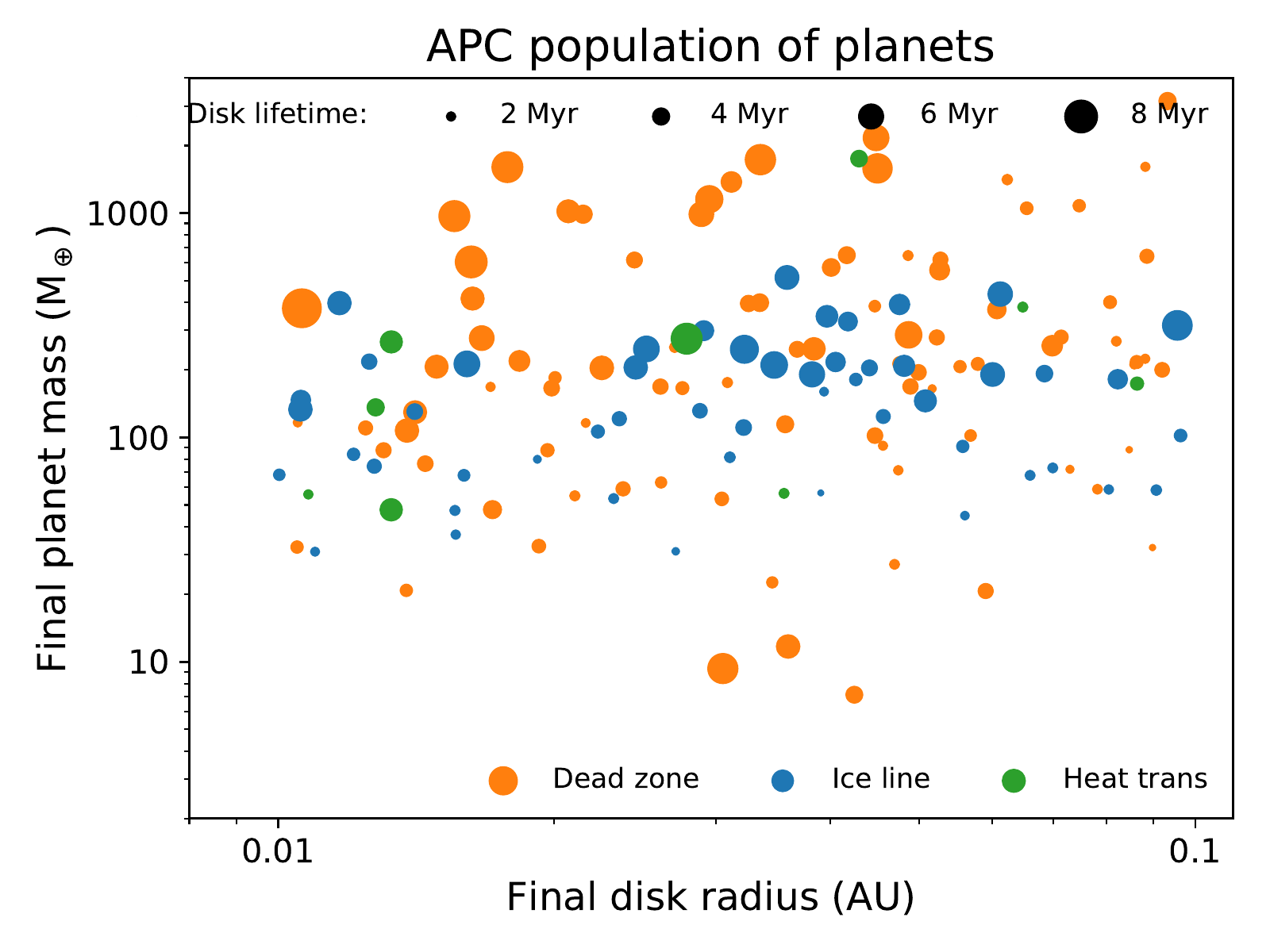}
\caption{The test population of Hot Jupiters and close in super-Earths taken from APC. This population covers a wide range of final radii and mass. We have noted whether the planet grew in the dead zone, water ice line, and the heat transition (heat trans) as well as the lifetime of the disk in which the planets grew.}
\label{fig:res01}
\end{figure}

In figure \ref{fig:res01} we show the final mass and orbital radius of the planets in the population from APC. A benefit of this population is that there is generally little bias, showing no structure in this parameter space. We do find that the lower mass range of the population are generated in disks with lower lifetimes. This is caused by the growth of the planet being cut off by the final evaporation of the disk gas (also see \citealt{AP18}).

On the other hand, the three lowest mass planets form in disks with the longest lifetime. This is caused by the fact that long lived disks tended to evolve slower (by construction, see eq. \ref{eq:19} with t$_{\rm dep} =$ t$_{\rm life}$). These light planets began at far radii in the dead zone in the most massive disks, and needed the full lifetime of the disk to migrate to their ending position. While massive, their natal disks tended to have lower metallicity, and hence their cores formation was less efficient. The location of the disk's dead zone is dominated by the surface density of dust and gas density which set the ionization level of the disk gas.

Planets with the largest mass and smallest orbital radii also occur in the longest lived disks, since they required the longest time to migrate inwards. These massive planets formed in heavier disks (and/or high metallicity disks) where they could grow more quickly then the lowest mass planets. These planets were all generated by proto-planets trapped at the dead zone (and one trapped at the heat transition). 

In what follows we compute the atmospheric C/O for this population of planets and search for any dependency on their planet formation history.

\subsection{ Ignoring solid accretion }\label{sec:nosoldidacc}

As a benchmark test we first compute the bulk chemical abundances of elements in the population when only the accretion of gas is considered.

\begin{figure}
\centering
\includegraphics[width=0.5\textwidth]{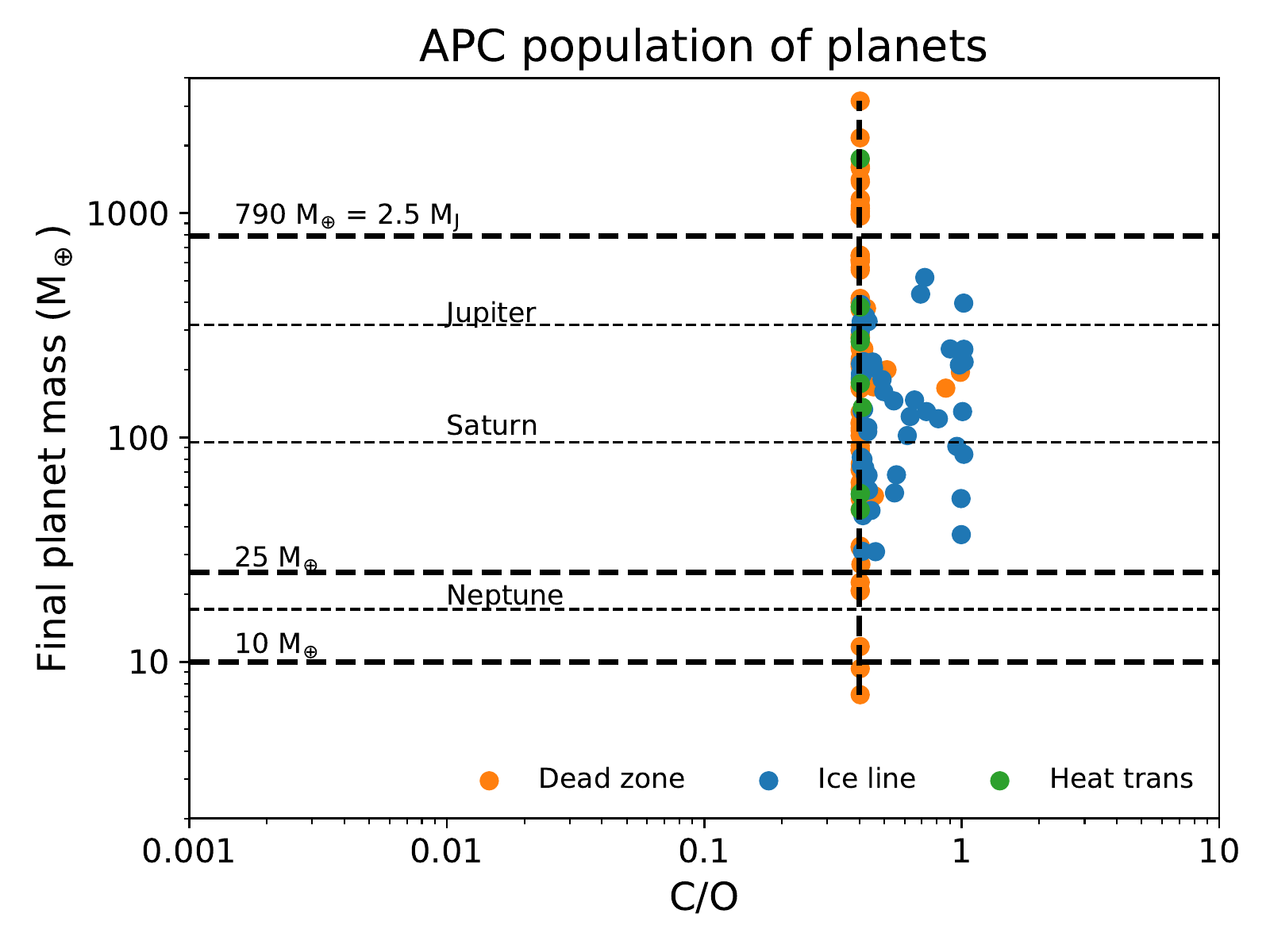}
\caption{Carbon-to-oxygen ratio for the population of planets assuming that gas accretion is the sole source of carbon and oxygen in the atmosphere. The vertical dashed line denotes the C/O of the gas disk, which the majority of planets mimic.}
\label{fig:res02}
\end{figure}

In figure \ref{fig:res02} we show the C/O for our population of planetary atmospheres. The vertical dashed line shows the bulk C/O of the disk gas. We find that the majority of planets result in a similar C/O as was initialized in the disk. This result is not surprising and has been seen in our previous work \citep{Crid17}. The main reason of this is that these planets have migrated inward of the water ice line before they accrete the majority of their gas. Inward of the ice line there are no volatiles frozen onto the grains, hence C/O remains unaffected with respect to the initial ratio.

For planetary masses between 25 M$_\oplus$ and 790 M$_\oplus$ we see our first deviation from the bulk C/O caused by the formation history of the planet. In this mass range C/O can get as high as $\sim 1$, owing to the fact that these planets have accreted their gas outward of the water ice line. Planets with C/O $>0.4$ grow in both the ice line and dead zone planet traps. 

Planets in the dead zone trap have simply not migrated within the ice line before they begin accreting their gas. Hence the gas is more carbon rich because water is in its ice phase. In the case of the ice line planets, their formation is occurring embedded within the dead zone of their disk (at smaller radii than the dead zone trap) - where the turbulent viscosity is much lower. As a result these planets easily open a gap, even when they are only a few times the mass of the Earth (recall equation \ref{eq:30}). They enter into a stage of Type-II migration much earlier than other protoplanets of that size, and lag behind the evolution of their ice line. This keeps these planets outward of the water ice line of the disk where C/O is larger than 0.4. In total we find 33.5\% of planets in our sample end up with larger C/O than the bulk disk gas.

\subsection{ Including solid accretion }\label{sec:yasolidacc}

While gas accretion surely dominates the mass of Jupiter-mass planets, solid accretion can contribute a significant fraction of the carbon and oxygen to their atmosphere. This comes from the fact that in the gas, carbon and oxygen make up approximately 0.01\% of the atoms, or about 0.1\% of the mass. Meanwhile the refractories are dominated by silicates (along with irons), and dust with ISM levels of carbon can be built up with a few tens of percent of carbon (by mass). Hence at least 100$\times$ more gas than solids must be accreted in order for the same number of carbon or oxygen atoms to be delivered to the atmosphere. This requirement becomes difficult for even a Saturn-mass planet (M $\sim$ 95 M$_\oplus$),

\subsubsection{Ignoring the carbon excess model}

\begin{figure}
\centering
\includegraphics[width=0.5\textwidth]{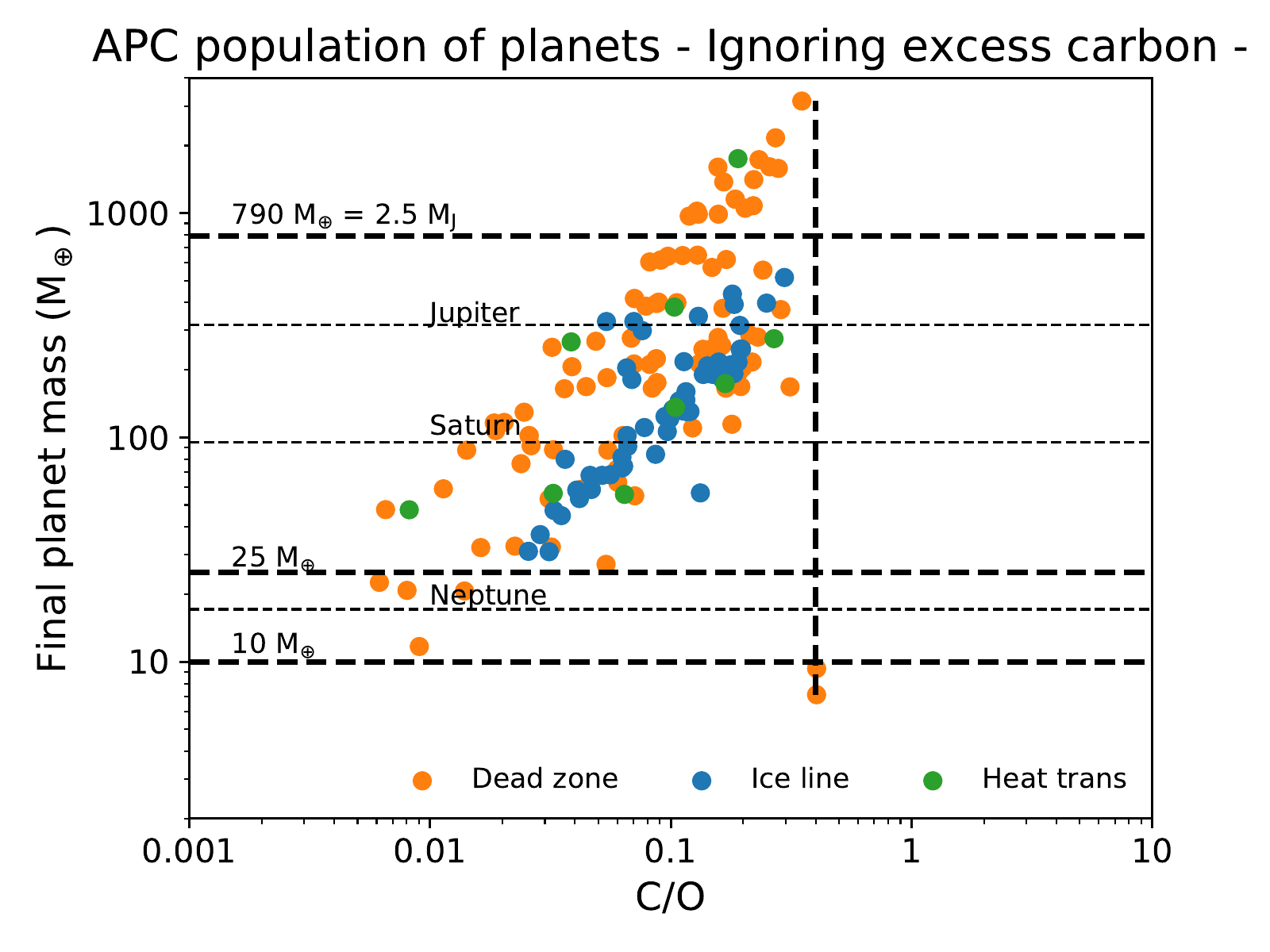}
\caption{Carbon-to-oxygen ratio for the case where we include solid accretion into the atmosphere,but ignore the effect of the excess carbon from refractory sources adding to the carbon budget. As such, solid accretion predominately delivers oxygen rich silicates and ices to the exoplanetary atmospheres.}
\label{fig:res03a}
\end{figure}

We begin in figure \ref{fig:res03a} by assuming that the excess carbon discussed in section \ref{sec:refc} does not contribute to the carbon content of the gaseous disk. In this way we ignore the impact that the depleted refractory carbon can have on the carbon budget of the disk gas. This is equivalent to assuming that the excess carbon moves with the bulk gas at a rate much faster (at least two orders of magnitude) than is assumed in the `reset' model. As such the only extra element available for accretion in this situation is the oxygen that is incorporated in the ice (if the planet is growing outside the water ice line) and silicates.

Without the extra carbon, we find in figure \ref{fig:res03a} that all of the planets end with smaller C/O than it did in the case of gas accretion alone. An exception to this can be found in our two lightest planets (with M $<10$ M$_\oplus$), which show no change in their C/O from the case of gas accretion alone. This arises because their atmosphere mass never exceeds 3 M$_\oplus$ which was the assumed cut off above which planetesimals will evaporate in the planetary atmosphere. Hence for these planets the incoming planetesimals reach the core of the planet, and only volatiles may contribute to the atmosphere (recall we ignore core erosion here). 

At slightly larger masses (10 M$_\oplus < $ M $<25$ M$_\oplus$) the atmosphere has surpassed the 3 M$_\oplus$ cut off, and refractory sources of carbon or oxygen can contribute to the atmosphere. For these planets the incoming carbon and oxygen is dominated by refractory sources, because the planet has not reached a large enough mass to undergo unstable gas accretion. As a result the ratio of gas and dust accretion timescales will be close to one and the accretion of refractory oxygen will be its dominant source.

\begin{figure}
\centering
\includegraphics[width=0.5\textwidth]{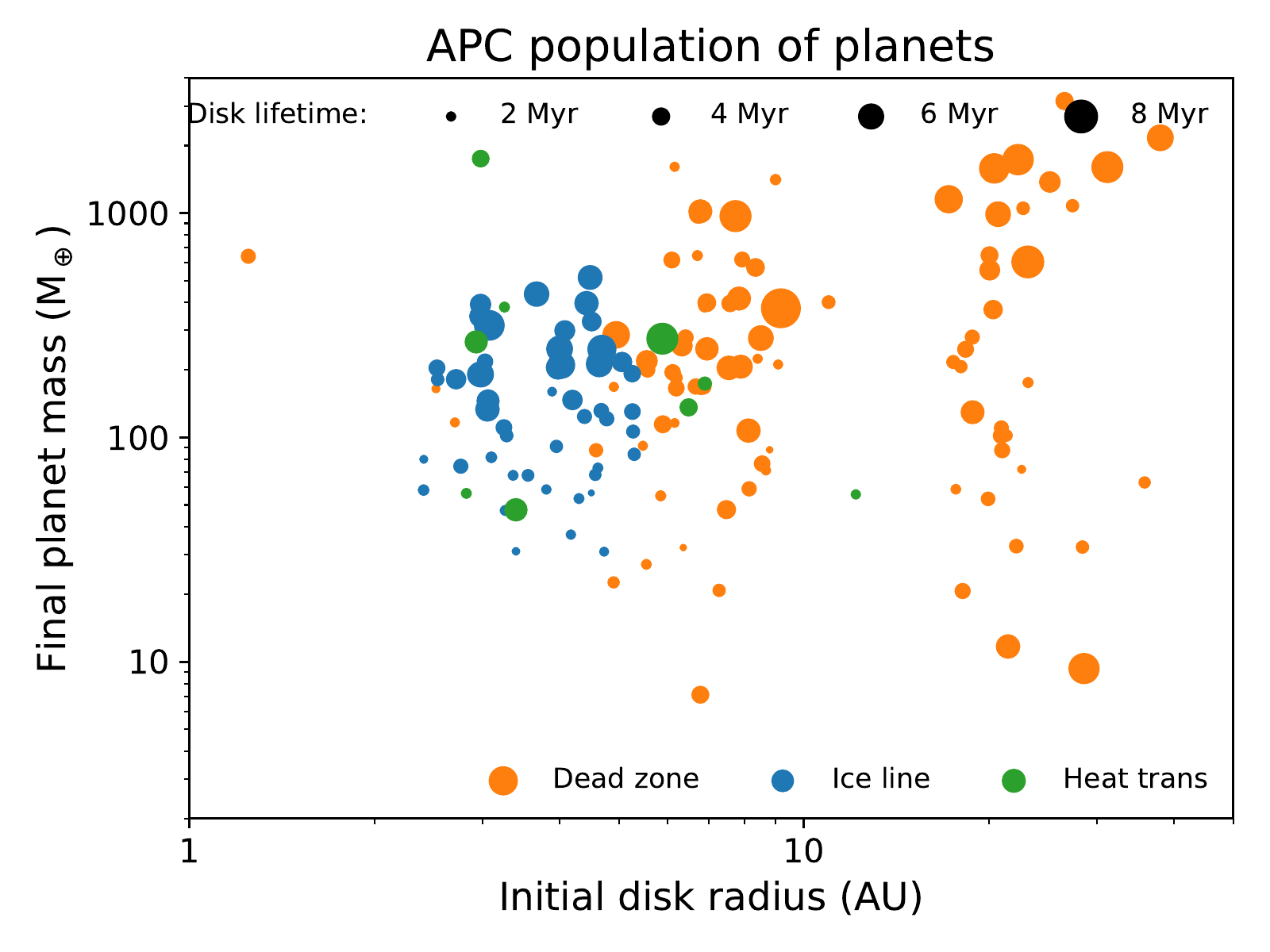}
\caption{Starting radii of the planets in our population. Generally the planets trapped at the ice line begin within a small range of radii since the location of the ice line is only weakly dependent on initial disk mass. Note that planets which start between 10 and 20 AU do not end their evolution as hot-Jupiter or close-in super-Earths, hence the deficit of points.}
\label{fig:res04}
\end{figure}

As for these lower mass planets, for masses $>$ 25 M$_\oplus$ the combination of solid and gas accretion determines the final C/O. We see a spread in final C/O depending on where the planet begins its evolution. For a given mass range, planets that were trapped at the water ice line (blue) or began in traps near the ice line tend to have higher C/O than planets that formed farther out in the disk. The planets which formed in the dead zone (orange) and heat transition (green) farther out than the water ice line (see figure \ref{fig:res04}), grew in a part of the disk where the surface density of solid material is reduced due to the radial drift of the dust (see for example \citealt{Crid16b}).

Since they start at further radii, the initial growth rate of planets at the dead zone and heat transition is slower than at the ice line (and in models where the dead zone and heat transition begin near the water ice line). These planets take much longer to build a core which is capable of undergoing unstable gas accretion, which allows more time for solids to accrete into the atmosphere. Ice line planets accreted much faster than planets starting in the dead zone or heat transition and hence accreted less solids (by mass) than planets which underwent slower core formation. In this model we lose planets with C/O $> 0.4$, since these planets also accrete (oxygen-rich) ices along with the silicates in the incoming planetesimals. 

\begin{figure}
\centering
\subfigure[ Final C/O for planets forming in the Reset model ]{\label{fig:res03b}
\includegraphics[width=0.48\textwidth]{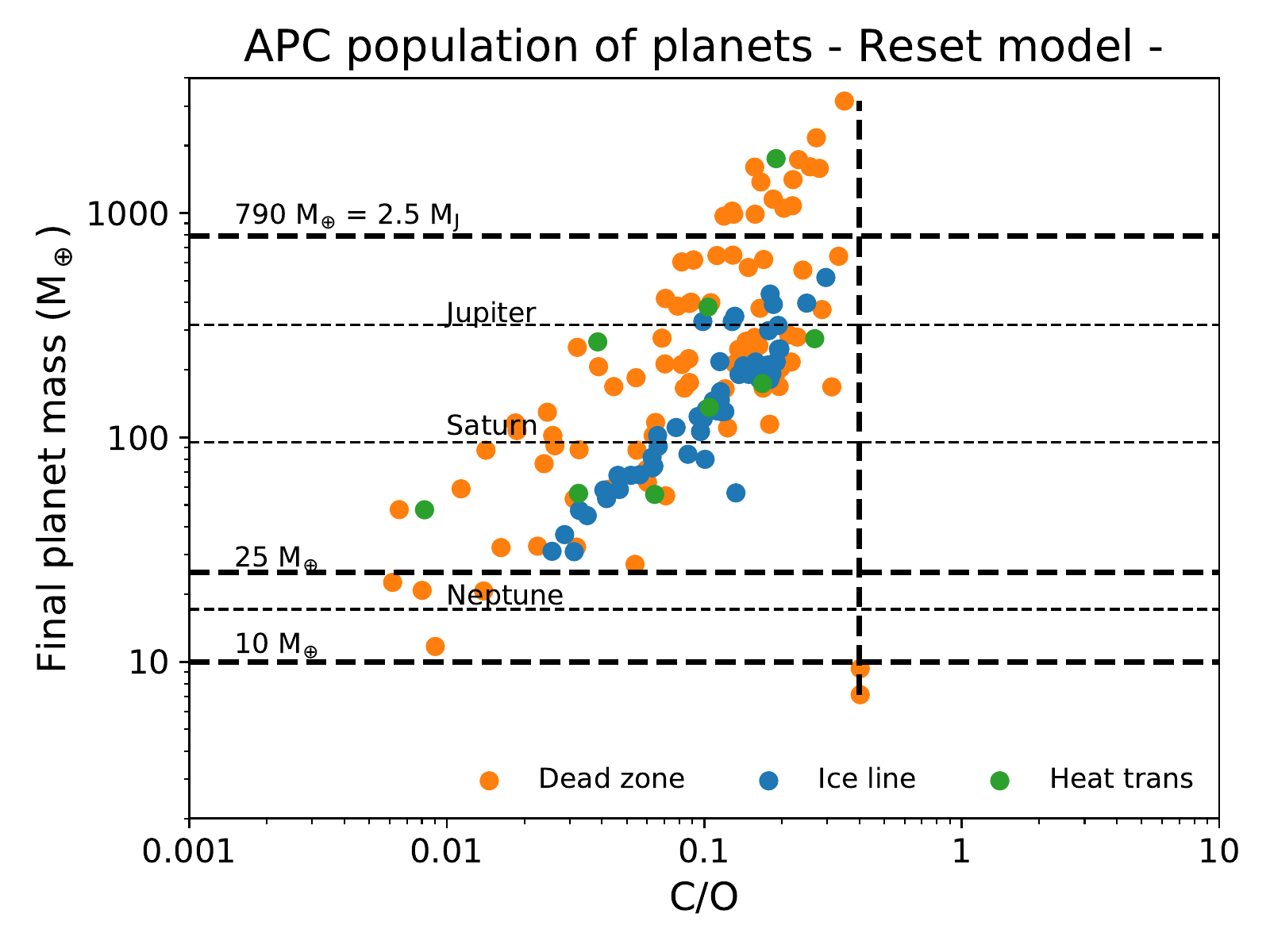}
}
\subfigure[ Final C/O for plnaets forming in the Ongoing model ]{\label{fig:res03c}
\includegraphics[width=0.48\textwidth]{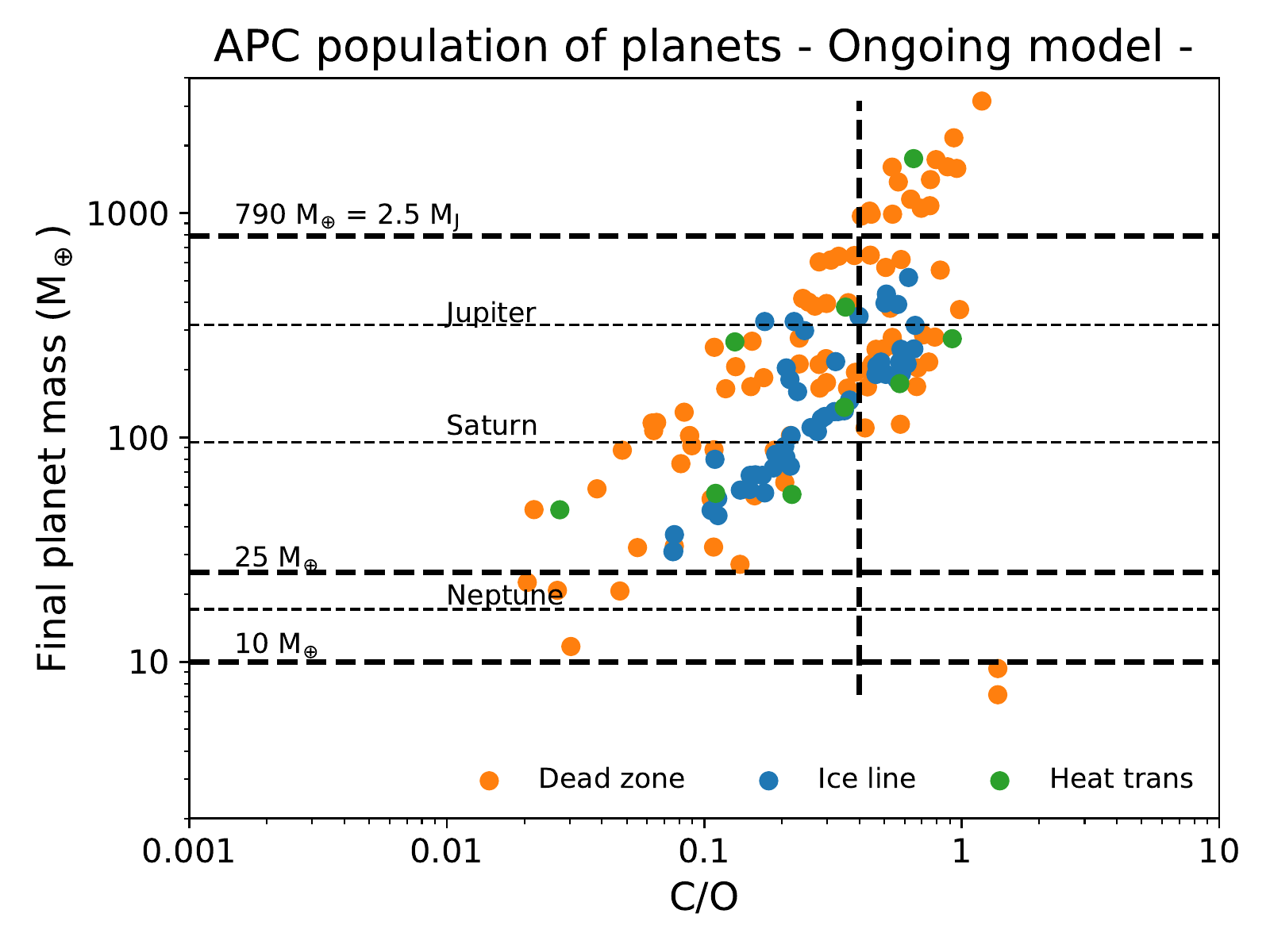}
}
\label{fig:res03}
\caption{Same as figure \ref{fig:res03a} but for the `Reset' model ({\bf Fig. \ref{fig:res03b}}) of carbon excess with a constant net advection speed of 6.32 AU Myr$^{-1}$ as prescribed by hydrodynamic disk models. As such, some planets reach a region of the disk where the excess carbon remains in the gas phase in time to incorporate it into its atmosphere. For the `Ongoing' model ({\bf Fig. \ref{fig:res03c}}) the net flux of the excess carbon is zero, which keeps a high carbon abundance. As such most of the planets reach a region of the disk where the carbon is enhanced due to the ongoing processing of dust grains.}
\end{figure}

\subsubsection{Including the carbon excess models}

In figures \ref{fig:res03b} and \ref{fig:res03c} we show the effect on the population from the two chemical models presented in \cite{Crid19b}. In the `reset' model (figure \ref{fig:res03b}) the excess carbon is allowed to advect with the bulk gas motion of the disk, quickly (in about $\sim 0.8$ Myr) accreting into the host star. Because of its rapid evolution, most of the planets are unaffected when moving from the model shown in figure \ref{fig:res03a} to figure \ref{fig:res03b}, however there are a few planets which evolve rapidly enough to be coincident with the region of the disk that has excess gaseous carbon when the planet undergoes gas accretion. 

For an example of this, two Jupiter-mass planets with C/O $< 0.1$ are shifted to C/O $\sim 0.2$ in the `reset' model of carbon excess. This is because these planets evolve fast enough that they accrete their gas in a region of the disk at an early enough time that the excess carbon has not disappeared. 

Finally in figure \ref{fig:res03c} we show the results of the `ongoing' model. Here the excess carbon is constantly replenished by the radial drift of dust grains, and hence the gaseous excess carbon remains high in the inner disk. When comparing to the model in figure \ref{fig:res03b} we find that all of the planets have shifted to the right (i.e. towards being more carbon-rich). 

This shift does return many of the higher mass (M $>$ 100 M$_\oplus$) planets to be more carbon-rich (C/O $>$ 0.4), however none of these planets result in atmospheres with C/O $>$ 1. The lowest mass planets (M $<$ 10 M$_\oplus$) also show carbon-rich atmospheres since they are not chemically effected by solid accretion, instead feeding exclusively on the carbon-rich gas present in this model.

\begin{figure}
\centering
\includegraphics[width=0.5\textwidth]{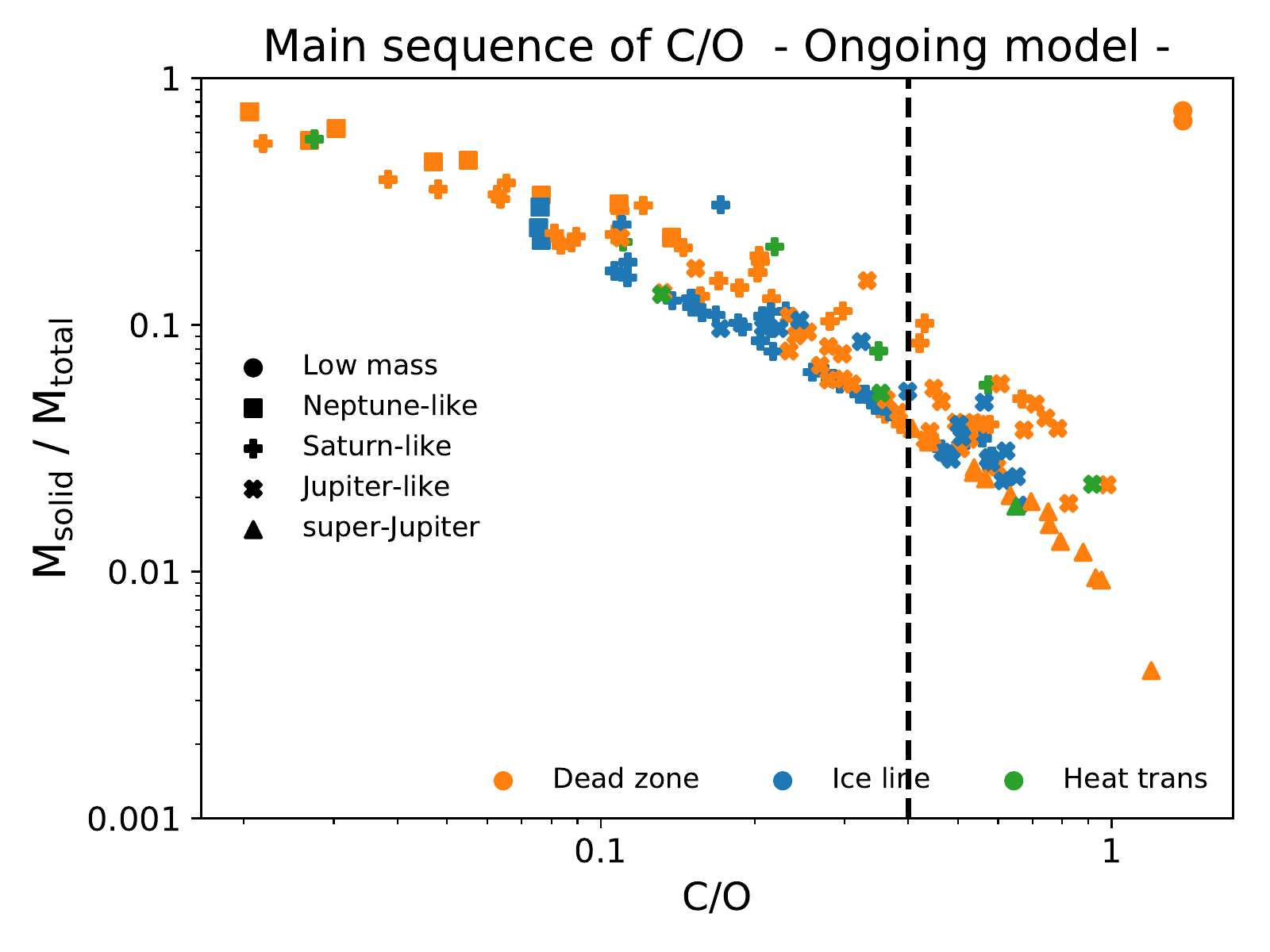}
\caption{C/O main sequence for the planet in our population. Over a wide range of masses we find that lower C/O is caused by the total mass of planet coming predominantly from solid (refractory) sources. The exception coming from the low-mass planets (M $<$ 10 M$_\oplus$), which are not chemically affected by the accretion of planetesimals.}
\label{fig:res05}
\end{figure}

\subsubsection{C/O main sequence}

In figure \ref{fig:res05} we generalize these results into a `main sequence' of C/O for the planets in this population. This sequence is relevant for planets that surpass the 3 M$_\oplus$ atmospheric mass limit to fully evaporate planetesimals in the atmosphere. We show here only the main sequence for the `ongoing' model since only minor differences (other than a horizontal shift) arises from changing the carbon refractory model. In the figure we have differentiated between low-mass (M $<$ 10 M$_\oplus$), Neptune-like (10 M$_\oplus$ $<$ M $<$ 40 M$_\oplus$), Saturn-like (40 M$_\oplus$ $<$ M $<$ 200 M$_\oplus$), Jupiter-like (200 M$_\oplus$ $<$ M $<$ 790 M$_\oplus$), and super-Jupiter (790 M$_\oplus$ $<$ M) planets. We see little spread off the main sequence from different mass ranges, except the low-mass planets which are far off the trend for their solid mass fraction. As was already said, this comes from the fact that these planets never have heavy enough atmospheres to evaporate incoming planetesimals.

For illustrative purposes, Jupiter has a solid to total mass ratio of $\lesssim 0.07$ \citep{Wahl2017}, according to figure \ref{fig:res05} results in a range of possible C/O between 0.2 - 0.6. Given that Jupiter's atmosphere is enhanced in carbon by a factor of 4 above the solar value \citep{Atreya2016}, then this range of C/O implies that its expected oxygen abundance should be between 10.6 - 3.6$\times$ the solar oxygen abundance. Of course, these simple calculations assume that Jupiter formed in a similar region as did the hot Jupiters in our model, near the water ice line. This requirement is consistent with the Grand Tack model of the formation of Jupiter and inner solar system, which starts Jupiter's outward migration inward of the water ice line.

In summary, the figure makes an important new point. For planets along the main sequence (that gain enough atmosphere mass to evaporate planetesimals) the smaller the contribution that solids have on the total mass of the planet, the higher their atmospheric C/O. The total mass contribution by solid accretion could additionally be interpreted as the atmosphere `metallicity', which we explore below.

\subsection{ Mass-metallicity relation }\label{sec:mmrel}

In our solar system, high mass planets tend to have lower metallicity \citep{Atreya2016}, which is often attributed to the amount of solid material that was accreted during their formation (relative to the gas accreted). More generally this trend has been inferred in the exoplanet population (see for example \citealt{MillerFortney2011,Thorngren2016,Thorngren2018}) as derived from interior structure modelling. 

\begin{figure}
\subfigure[ Mass-metallicity relation measured with O/H ]{ \includegraphics[width=0.48\textwidth]{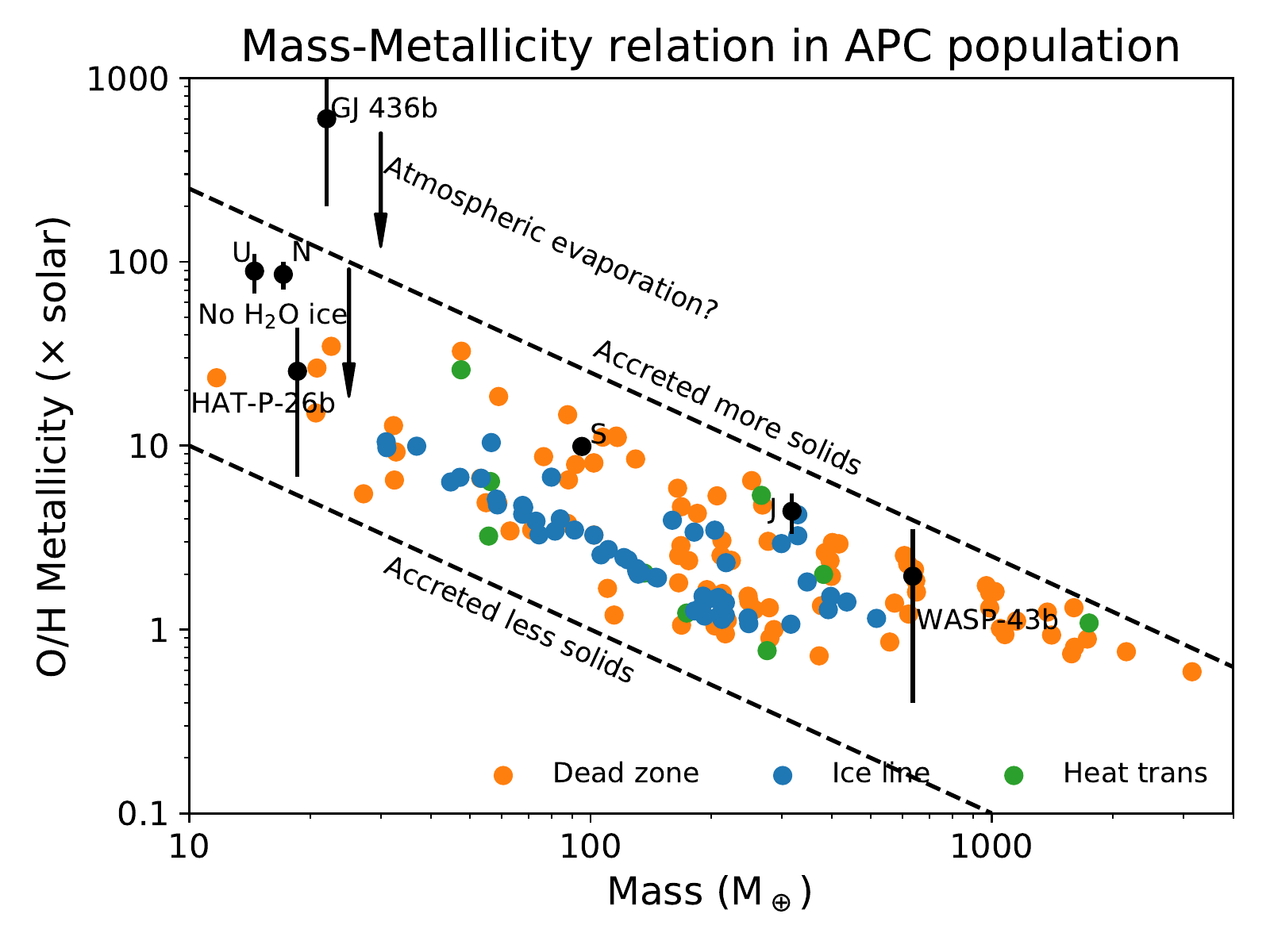}
\label{fig:massmetal01}
}
\subfigure[ Mass-metallicity relation measured with Si/H ]{ \includegraphics[width=0.48\textwidth]{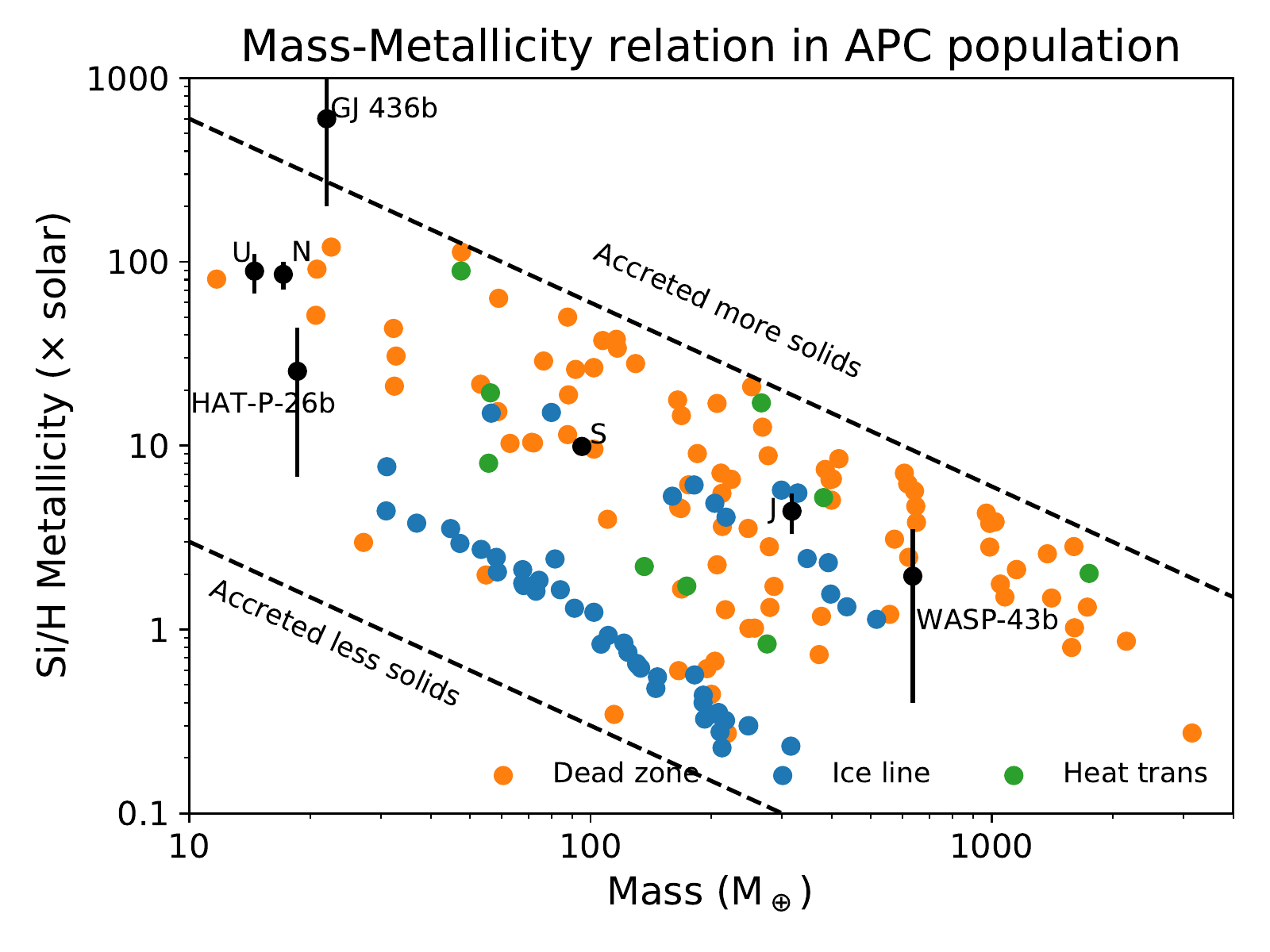}
\label{fig:massmetal02}
}
\caption{Mass-metallicity relation for two tracers, the oxygen-to-hydrogen ratio ({\bf Fig. \ref{fig:massmetal01}}) as well as the silicon-to-hydrogen ratio ({\bf Fig. \ref{fig:massmetal02}}). We similarly show the metallicity for the solar system giants (inferred by methane abundance, and taken from \citealt{Kre14}) as well as WASP-43 b \citep{Kre14}, GJ 436 b \citep{Morley2017}, and HAT-P-26 b \citep{MacDonald2019} from their water abundance. For a given mass, lower metallicity planets generally came from the water ice line, accreting very quickly which limited the amount of solids that could be accreted into the atmosphere. We find that our Neptune-mass planets are consistent with the metallicity of HAT-P-26 b, but not with Uranus or Neptune, suggesting the formation location could be important to O/H. The Si/H spread in metallicity for a given planet mass is caused by the accretion of solids. In this case our Neptune-mass planets are consistent with the inferred metallicity of Uranus and Neptune.}
\end{figure}

In figures \ref{fig:massmetal01} and \ref{fig:massmetal02} we show the mass-metallicity relation for our population of planets given two methods of inferring the metallicity: oxygen-to-hydrogen (O/H) and silicon-to-hydrogen ratio (Si/H). We compared our population of planets to the solar system giants and three exoplanets. The range of metallicity for the solar system giants (inferred by methane abundances) and WASP-43 b (inferred from water abundance) were taken from \cite{Kre14}. The metallicity of the hot-Neptunes GJ 436 b and HAT-P-26 b were inferred by their water abundance by \cite{Morley2017} and \cite{MacDonald2019} respectively.

In figure \ref{fig:massmetal01} we show the atmospheric metallicity (as inferred by O/H) for our population of planets. In this way we can best compare to the exoplanet observations, since their metallicity was inferred by an abundant oxygen carrier. Here we see that HAT-P-26 b and WASP-43 b agree well with O/H of their respective mass range. In the Neptune-mass range we under-predict the O/H metallicity of Uranus and Neptune, which can be explained by a lack of icy planetesimal accretion in HAT-P-26 b while the opposite is true for Uranus and Neptune. This suggests that the Neptune-like planet HAT-P-26 b likely formed inward of its disk's water ice line.

For GJ 436 b we see that we have under-predicted its metallicity by almost two orders of magnitude. Since it is known to orbit very close to its host star this can be explained by a significant amount of atmospheric evaporation.

In figure \ref{fig:massmetal02} we compare the metallicity of each of our planets as inferred by Si/H to the metallicity of the solar system giants and three exoplanets. Once again we see that the relation between planet mass and metallicity inferred by \cite{Kre14} agrees well with the general trend in the figure. The large spread is caused by the amount of solid accretion that the growing planets have experienced. The spread is larger than we find in figure \ref{fig:massmetal01} because the source of silicon is strictly from planetesimal accretion, while oxygen can also be accreted directly from the gas.

Generally we find that the mass-metallicity relation is most determined by the fraction of solid mass accretion relative to the total mass of the planet. However different tracers of metallicity can be subject to factors other than the total mass accretion. The O/H can be most sensitive to variations due to the position of the disk relative to the water ice line, as well as the physical history of the atmosphere after formation. In particular, Neptune-mass planets are most sensitive to their formation history because a majority of their oxygen can come from refractory sources.

We note that along with our C/O main sequence introduced in the previous section, the mass-metallicity implies an anti-correlation between planet metallicity and atmospheric C/O, as discussed by \cite{Madhu14,Madhu16PA,Espinoza2017}. In our work this anti-correlation is a direct result of the relative importance that solid accretion has on the total mass of the planet, and the quantity of solids that can be delivered into the planetary atmosphere. These solids are oxygen-rich (relative to carbon) as proposed by \cite{Oberg11} when we include refractory carbon erosion, hence Neptune-like planets that are dominated by solid accretion are found to have remarkably low C/O.

\section{ Discussion: What does observed C/O say about formation history? }\label{sec:discussion}

A key question we wish to answer is whether a measured C/O can give any insight into the formation history of a planet. As we have seen, this interpretation is complicated by the treatment of solid accretion, and the excess carbon expected to be generated from refractory sources.

\subsection{ General trends }\label{sec:trends}

Generally we see that the low mass (M $<$ 10 M$_{\oplus}$) planets are not affected by solid accretion, because their atmosphere are not heavy enough to evaporate incoming planetesimals. Additionally our model suggests that these close in, low mass planets have accreted their gas within the water ice line. These planets could give us a secondary method of probing refractory carbon erosion. However they are difficult to characterize due to their size, and will likely require the {\it James Webb Space Telescope} (JWST) to efficiently probe. A second possibility for these low mass planets exists: they grow outside of the ice line, and are dynamically scattered into the inner solar system after the evaporation of the disk gas - hence the presence of an unknown companion may be inferred. Both this, and details of carbon refractory erosion require follow up observations to confirm, but would nevertheless add to our ever growing knowledge of planet formation. 

At slightly higher mass ranges (10 M$_\oplus$ $<$ M $<$ 25 M$_\oplus$, squares in figure \ref{fig:res05}) the accretion of solids will dominate the delivery of carbon and oxygen. In the case of close-in planets, our models predict that this mass range will be highly oxygen rich, owing to the large fraction of mass being composed of silicates. For planets in much longer orbits than investigated here (as in Neptune) one might expect much higher C/O ratios due to the accretion of more carbon rich planetesimals, along with enrichments of both O/H and C/H - as is observed with Neptune \citep{Atreya2016}. Even in the case that excess gaseous carbon is included (figure \ref{fig:res03c}) these planets remain at very low C/O. Hence we expect very strong water emission / absorption features to be present in these types of close-in planets. Because of their dependence on the solid accretion history, these Neptune-like planets could act as a useful tracer of alternative accretion models like pebble accretion \citep{Johansen2007,Ormel2010,Bitsch2015}.

For Saturn- and Jupiter-like planets (25 M$_\oplus$ $<$ M $<$ 790 M$_\oplus$, diamonds and stars in figure \ref{fig:res05}) we find a larger deviation in the bulk C/O which depends on where the planet began accreting. For planets beginning near the water ice line (trapped in any of the three traps), their bulk atmospheric C/O stays within an order of magnitude of the fiducial gas disk C/O. However this is heavily dependent on the excess carbon model

Planets that began their evolution outside the water ice line generally have accretion timescales that are long due to the lower dust (and hence planetesimal) column density, and hence can accrete a large fraction of their carbon and oxygen from refractory sources. Hence sub-Saturn mass planets (M $< 100$ M$_\oplus$) should have sub-stellar C/O while larger planets may accrete enough gaseous carbon to have stellar-like (or larger) C/O.

For the largest planets (M $>$ 790 M$_\oplus$, triangles in figure \ref{fig:res05}) the effect of solid accretion results in only small scatter, since the majority of their mass comes directly from gas accretion. The most massive planets have the highest C/O since their gas-to-solid accretion is the highest. They are very dependent on which model is controlling the distribution of the carbon excess. In the case that the excess carbon from the refractories disappears due to it flowing in with the bulk disk gas (i.e. the `reset' model), all of the highest mass planets have sub-stellar C/O. 

To have atmospheres with C/O$_{\rm planet}>$ C/O$_{\rm star}$ we require an ongoing process producing excess carbon in the gas phase that is accreted by these growing planets. The shift to the right between figures \ref{fig:res03b} and \ref{fig:res03c} (also see figure \ref{fig:diss01}) depend on what extent the refractory carbon is depleted from the grains in the region where the planet formed. Hence in the case that excess carbon survives for the full lifetime of the disk, then all of the highest mass planets have stellar-like or super-stellar C/O - some even approaching C/O $\sim 1$. This suggests that if the detection of high C/O among massive planets continues to be common, the `ongoing' model might be the best description of the excess carbon distribution.

\subsection{ Comparing to observations }\label{sec:compare}

As we implied earlier, comparing our synthetic chemical population of planets to the observed population is necessary for understanding the properties of planet formation.

\ignore{
\begin{figure}
\centering
\begin{overpic}[width=0.5\textwidth]{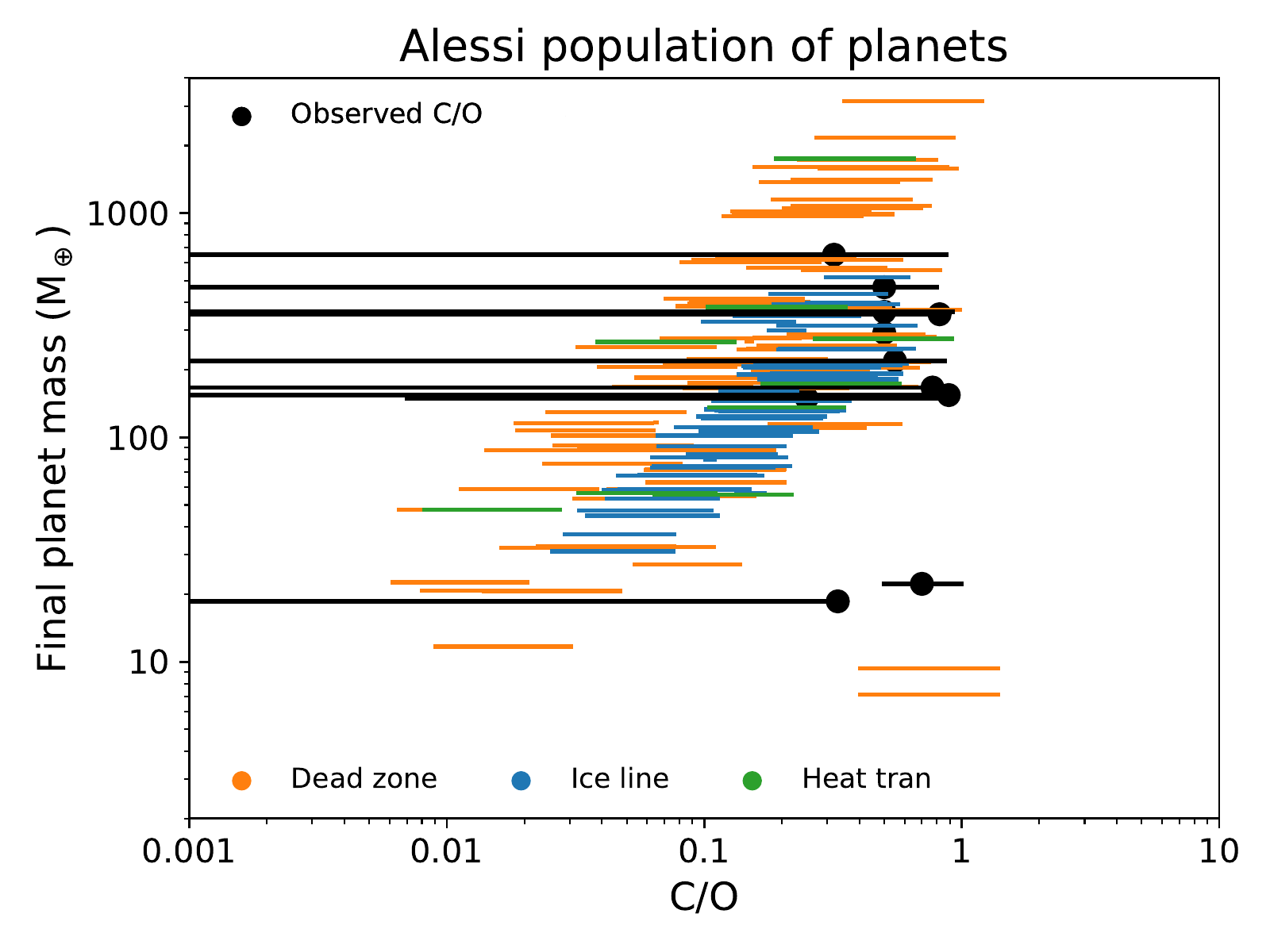}
\put(75,32){ GJ 436 b }
\put(64,24){ HAT-P-26 b }
\end{overpic}
\caption{Comparing the observed atmospheric C/O shown in table \ref{tab:diss01} to our population of planets. Here each line segment represents the shift in C/O between the carbon excess models shown in figure \ref{fig:res03b} and \ref{fig:res03c} for each synthetic planet. }
\label{fig:diss01}
\end{figure}
}

In figure \ref{fig:diss01} we show a comparison between our synthetic population of planets to a small set of characterized planetary atmospheres. The individual planets, their values / ranges, and references are found in table \ref{tab:diss01}. The coloured line segments note the shift in C/O caused by the two carbon excess models shown in figures \ref{fig:res03b} and \ref{fig:res03c}. The points mark a specific C/O outlined by observation while the black line segments denote possible ranges of observed planets determined through retrieval methods. 

\begin{table*}
\centering
\caption{List of observed C/O and/or inferred ranges of possible C/O for planets that orbit at $< 0.1$ AU. Where applicable, C/O were estimated based on the published abundances of observed molecules. Cases where C/O $=0.5$ come from best fit models where the abundance of CO and H$_2$O are equal. }
\label{tab:diss01}
\begin{tabular}{| c | c | c | l | c | l |}
\hline
Planet & mass (M$_\oplus$) & C/O & Reference & C/O range & Reference \\\hline
51 Peg b & 149 & 0.25 & \cite{Brogi2013} & $0.007 <$ C/O $ < 0.95$ & \cite{Birkby2017} \\\hline
GJ 436 b & 22 & 0.7 & \cite{Morley2017} & $0.5 <$ C/O $<1$ & \cite{Madu2011} \\\hline
HD 179949 b & 292 & 0.5 & \cite{Brogi2014} & & \\\hline
HD 189733 b & 363 & 0.5 & \cite{Brogi2016} & C/O $<0.92$ & \cite{Benneke2015} \\\hline
HD 209458 b & 219 & 0.5 & \cite{Rimmer2016} &  C/O $ < 0.86$ & \cite{Benneke2015} \\\hline
WASP-12 b & 467 & 0.5 & \cite{Stevenson2014} & C/O $< 0.80$ & \cite{Benneke2015} \\\hline
WASP-43 b & 652 & 0.319 & \cite{Feng2016} & C/O $< 0.87$ & \cite{Benneke2015} \\\hline
HAT-P-1 b & 167 &  & & C/O $< 0.77$ & \cite{Benneke2015} \\\hline
HAT-P-26 b & 19 & & & C/O $< 0.33$ & \cite{MacDonald2019} \\\hline
WASP-17 b & 154 & & & C/O $< 0.89$ & \cite{Benneke2015} \\\hline
WASP-19 b & 354 & & & C/O $< 0.82$ & \cite{Benneke2015} \\\hline
\end{tabular}
\end{table*}

Generally speaking these planets all lie above C/O $= 0.4$, placing them in the population of planets that begin their growth either trapped at the water ice line, or at a planet trap that originated near the ice line. This tendency of beginning their evolution near the water ice line is consistent with the works of \cite{Drazkowska2017} which suggest that a dust pile up at the water ice line due to radial drift leads to efficient planetesimal formation which should then lead to more rapid growth of the first planetary cores.

A particular oddity appears to be in the hot-Neptune GJ 436 b (as reported by \citealt{Madu2011} and \citealt{Morley2017}) which appears near the bottom right part of figure \ref{fig:diss01}. With its mass of $\sim 22$ M$_\oplus$ it is in a mass regime that we attribute with predominately (very) low C/O, owing to a large fraction of its accreted mass being from solid bodies. This {\it could} be a sign that the excess carbon from refractory carbon erosion is not a universal process, since a system that has not undergone refractory carbon erosion would produce carbon-rich Neptune-mass planets at small radii from its carbon-rich (i.e. ISM-like) refractory component.

This discrepancy can also be explained by chemical processing of the gas within the atmosphere over the following billions of years after formation. With the silicon that is delivered to the atmosphere, some of the oxygen can be locked back into refractory material that precipitates out of the upper layers of the atmosphere. This would increase the observed C/O as described in \cite{Helling14}, assuming that the carbon is not also captured in these refractory grains.

\begin{figure}
\centering
\begin{overpic}[width=0.5\textwidth]{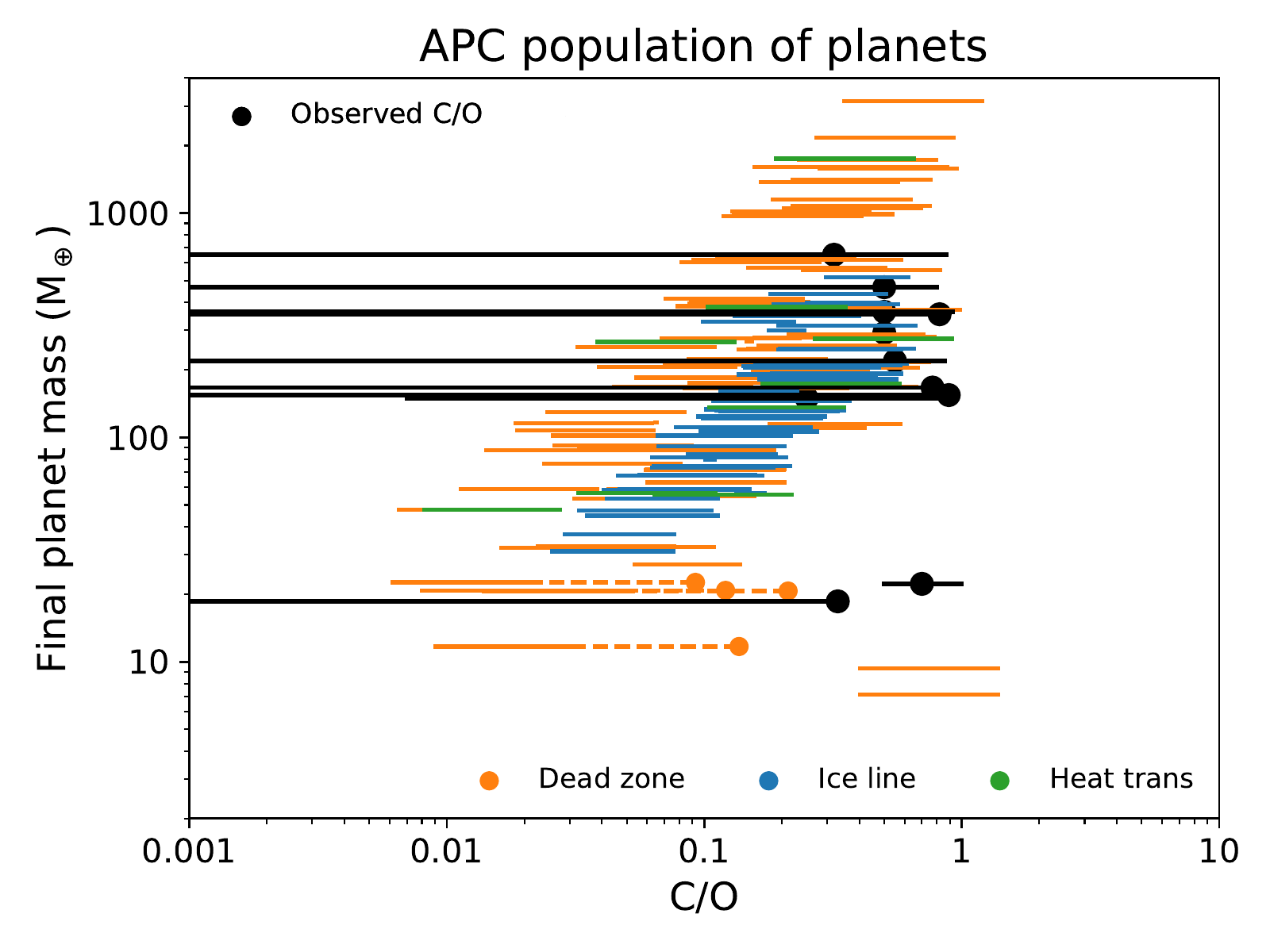}
\put(75,32){ GJ 436 b }
\put(64,24){ HAT-P-26 b }
\end{overpic}
\caption{Comparing the observed atmospheric C/O (black) shown in table \ref{tab:diss01} to our population of planets. Here each line segment represents the shift in C/O between the carbon excess models shown in figure \ref{fig:res03b} and \ref{fig:res03c} for each synthetic planet. Additionally we include an increase in C/O in Neptune / GJ 436 b -like planets caused by a rain out of silicates (dashed line, see the text). }
\label{fig:diss01}
\end{figure}

To that end we run a simple test of chemical equilibrium models at a range of temperatures (500 K - 1500 K) and gas pressures ($10^{-4}$ bar - 0.1 bar) with the initial elemental abundances of H, C, O, N, and Si as computed by our planet formation model for the four planets with masses near to GJ 436 b. We find that the observable C/O (i.e. in CO, CO$_2$, CH$_4$ and H$_2$O) increases by roughly a factor of 4.5 (across the range of pressure and temperatures used here) because some of the oxygen condenses out of the gas in silicates and (presumably) rains out of the upper atmosphere or produces clouds.

In figure \ref{fig:diss01} we also show the shifted C/O for the four Neptune-mass planets by the factor of 4.5 from our chemical equilibrium model. Clearly the increased C/O shift these planets to the right of the figure, however not far enough to explain GJ 436 b. Hence chemical processing after the atmosphere has formed is insufficient to fully explain this planet.

A second planet with a similar mass is HAT-P-26 b which has a recently reported C/O upper limit by \cite{MacDonald2019} of 0.33. It is more consistent with our maximum allowed C/O from our formation model and equilibrium chemistry. \cite{MacDonald2019} also report the detections of metal hydrides (TiH, CrH, and ScH) at lower ($< 3.6\sigma$) signal to noise. These gas phase detections at the moderate ($\sim 500$ K) temperature of HAT-P-26 b suggest that non-equilibrium processes must be dominating over pure chemistry (as assumed in our chemical equilibrium model). The complexity of the chemical and physical processing of atmospheres in this mass range is intriguing, but we leave it to future work.

Currently most observed C/O (including upper limits) in exoplanetary atmospheres fall within an order of magnitude. This is partly caused by the fact that current observations do not cover a wavelength range where the most abundant carbon carriers (CO, CO$_2$) have strong spectral features. This will be greatly improved over the next decade with the flight of JWST, and the wide wavelength range of its instruments. 

In summary the observed atmospheric C/O all fall within an order of magnitude of each other, regardless of planetary mass. This suggests that the amount of post-formation chemical processing in the atmosphere will be more pronounced for the lower mass planets, since their fraction of oxygen in the atmosphere coming from refractory sources is higher. As such high mass planets do not undergo as much chemical processing as discussed above since their Si/O is lower. Additionally these Neptune-like planets could be more sensitive to chemical interactions between the atmosphere and core, which we do not account for in this model.
 
\subsection{Comparison with previous theoretical work}

As previously mentioned, other theoretical studies have struggled to produce high super-solar C/O due to their simplified chemical model \citep{Mousis2011,AliDib2017}, or depleted gas disk caused by the inward accreting bulk gas \citep{Thiabaud2015,Mordasini16}. In these cases the accretion of solids leads to very low C/O due to predominately oxygen rich building blocks (both planetesimals and pebbles). Because of this tendency, explaining some recent observations which suggest C/O has been difficult to reconcile.

In the case of \cite{AliDib2017}, a high C/O was achieved by assuming a large core erosion efficiency, which incorporated the carbon and oxygen from the metal rich core into the atmosphere. Their refractory component incorporated an assumed quantity of 80\% and 50\% of the total carbon and oxygen respectively. In our model, refractory carbon and oxygen represents 70\% and 40\% respectively, resulting in a higher refractory C/O (=1.4) than theirs (=0.88). In this case, the impact of core erosion could be more strong than is suggested in \cite{AliDib2017}, however numerical simulations have shown that the mixing efficiency from the core to the atmosphere can be more inefficient than previously thought \citep{Moll2017}.

\cite{Thiabaud2015} and \cite{Mordasini16} deplete their volatile carbon and oxygen from the gas by accreting it into the host star within the first Myr. The gas that replaces the accreted gas comes from farther out in the disk where the volatiles have frozen out. While this gas moves inward, the frozen volatiles would similarly move inward, where they would also lose their ices to desorption \citep{Bosman2017b,Booth2019}. Hence the volatile component of the disk is maintained over a longer timescale where it can be available for accretion onto growing planets. Similarly, extra carbon and oxygen can be delivered just within the ice lines of various volatiles - potentially offering a separate pathway of enhancing the carbon and oxygen component of atmospheres \citep{Booth2019}.

Moreover, the studies which include the impact of refractory carbon erosion in the composition of the refractories have not included the enhanced carbon in the gas that would result. In this way their disks tended to be very oxygen rich in the inner solar system, causing the resulting exoplanetary atmospheres to be similarly oxygen rich. Here we have used the excess carbon models of \cite{Crid19b} to account for the abundance and evolution of the carbon released by the erosion of refractory carbon. In this way the gas can become quite carbon rich (recall Figure \ref{fig:ctoomap}) in the region of the disk where the majority of planets accrete their gas.

In our work we can reproduce high C/O in planets with high mass (M$_{\rm plnt} > 100$ M$_\oplus$) only if refractory carbon is continuously eroded throughout the lifetime of the disk. If this is not the case, our models revert to previous studies which predict solar or sub-solar C/O. We have not considered the erosion of the planetary core in this work, however its role in setting the final atmospheric abundances could be important. Similar to aforementioned studies we do not produce C/O $>$ 1, except for planets with mass $<$ 10 M$_\oplus$ in the `ongoing' carbon excess model. This result is simply because the chemical composition of these atmospheres is not impacted by the accretion of planetesimals, and hence only impacted by gas accretion.

\subsection{Caveats}\label{caveats}

In principle, there are carbon rich planetesimals outside of the 5 AU radius which we have set as the radial extent of the carbon erosion region of the protoplanetary disk. Indeed comets that originate from orbits farther out than Jupiter do show ISM levels of C/Si in our own solar system \citep{Berg15}. However every planet in our population begins accreting gas inside of 5 AU, hence any planetesimal that it can accrete (according to our model) will be carbon poor. This does ignore two possible complications to our model: firstly the core of the planetesimal can be built outside of the 5 AU extent of the carbon erosion region. The carbon rich core could  release its carbon into the atmosphere by out gassing or core erosion as in \cite{Madu2014}. This effect is a complicated physical problem, and hence we have ignored its contribution to the carbon content completely. However this effect could be important for Neptune-like planets where the core makes up a higher percentage of the total planetary mass.

Second we have ignored the possible effect of dynamical mixing of carbon-rich planetesimals from the outer disk into the inner disk, which could potentially deliver extra carbon to growing planets in the inner disk (r $<5$ AU). This process apparently did not happen in our solar system (since carbon erosion can be seen today) and could be caused by the presence of Jupiter acting as a guard to the inflow of carbon rich solids. Systems that lack a Jupiter-like planet at radii outside the 5 AU erosion region, could continue to have refractory carbon delivered to the inner disk, and if there is no `ongoing' process actively taking carbon away from the solids (as is assumed in the `reset' model) then extra refractory carbon would be available for accretion.

Finally, Jupiter should have acted as a barrier to inward drifting dust grains in our own solar system. As such it would have stopped the flow of incoming refractory carbon in the ongoing model. In a more general case, planetary systems that also contain a Jupiter-like planet at larger radii could see an ongoing-like model of refractory carbon erosion being suppressed. Given the importance of the ongoing model to explain the observed C/O of characterized exoplanetary atmospheres, more work is needed to understand the details involved in depleting the refractory carbon in inner planetary systems.

\section{ Conclusions }\label{sec:conclusion}

In this work we have combined the method of planet population synthesis with the astrochemistry of protoplanetary disks to predict the atmospheric C/O in a population of hot-Jupiter and close-in super-Earth planets. For gas accretion alone we find that the majority of planetary atmospheres have C/O that match the elemental ratio of the gaseous protoplanetary disk. However for a small subset of planets we find C/O that exceed the protoplanetary disk gas, due to the fact that they accreted their gas outward of the water ice line where water has frozen out on the dust grains.

Going beyond simple gas accretion we include the chemical effect of accreting planetesimals into the growing planetary envelope. We simply parametrize detailed studies of planetesimal survival through an atmosphere, and assume that if the gaseous envelope is below 3 M$_\oplus$ then the refractory material in the planetesimal reaches the core. Otherwise the planetesimal is completely destroyed in the atmosphere. In both cases, we assume that the volatile component of the planetesimal is delivered into the atmosphere.

Here the consequence of solid accretion on the amount of carbon and oxygen in the atmosphere depends on the fraction of accreted mass that comes from planetesimals, and whether those planetesimals deliver their material into the atmosphere or to the core. The chemical make-up of these planetesimals includes the effect of refractory carbon erosion in our model, and hence within 5 AU the majority of refractories are oxygen rich. Broadly speaking our results show:\begin{itemize}
\item Low mass (M $<$ 10 M$_\oplus$) planets do not accrete enough of an envelope to fully destroy planetesimals as they pass through the growing atmosphere. C/O is determined by gas accretion alone.
\item Neptune-mass (10 M$_\oplus$ $<$ M $<$ 25 M$_\oplus$) planets have enough of an envelope to evaporate incoming planetesimals before they reach the core. They accrete a high fraction of their carbon and oxygen from these refractory sources, and hence are oxygen rich.
\item Saturn- and Jupiter-mass planets (25 M$_\oplus$ $<$ M $<$ 790 M$_\oplus$) show a wide range of atmospheric C/O due partly to the fraction of refractory sources that contribute to the carbon and oxygen. Additionally this mass range shows some deviation in where planets begin their evolution. Planets that start either in the water ice line, or near the water ice line (but in another trap), tend to have higher C/O than planets that begin at larger orbital radii.
\item Super-Jupiter-mass (M $>$ 790 M$_\oplus$) planets show less of a tendency of starting their growth near the water ice line. They follow a similar trend as the Jupiter-mass planets which started their evolution outward of the water ice line.
\item All planets above a mass of 10 M$_\oplus$ fall along a `main sequence' between C/O and the fraction of solid mass that make up the total mass. This main sequence applies for planets that have accreted enough of a proto-atmosphere that incoming planetesimals evaporate completely.
\item The mass-metallicity relation is explained well with the fraction of solid mass (relative to total mass) that is accreted into the atmosphere. However variations from formation history can be particularly important to the inferred metallicity of Neptune-mass planets if the inferred metallicity is from O/H.
\end{itemize}

We compared our results to the inferred C/O from a number of planets who's atmospheres have been chemically characterized, and who's orbital radius is less than 0.1 AU. We find that:\begin{itemize}
\item Observed hot-Jupiters are consistent with the higher range of C/O found in our synthetic planets. This tendency suggests that these planets begin their accretion near the water ice line of their respective disks.
\item Neptune-like planets GJ 436 b and HAT-P-26 b have much higher C/O (by about two orders of magnitude) than inferred by our model. We show that part of this discrepancy could be caused by chemical processing which produces condensible materials which locks up oxygen and removes it from the gas in the upper atmosphere.
\item A discrepancy in low-mass C/O could also suggest that refractory carbon erosion is not a universal process, and the two planets we have included here were formed in a disk with carbon rich refractories. This could similarly suggest that refractory carbon can reform on the grains, as proposed by \cite{Wei2019}.
\item Additionally the discrepancy could be caused by the dynamical delivery of carbon rich planetesimals from the outer part of the disk. This effect however, is beyond the scope of our work here.
\end{itemize}

We have shown that variation of the atmospheric C/O ratios of planetary atmospheres do arise through a planet's mass accretion alone. We have seen this variation even though we have assumed that the stellar C/O does not vary between stellar systems. However as shown by \cite{Brewer2016b}, higher metallicity stars tend (with scatter) to have higher C/O. This suggests that some scatter in atmospheric C/O comes from the difference in the initial elemental abundances of the stellar system. To what extent this effect increases the scatter above planet formation alone will be an interesting line of research that we will explore in a future paper. In conjunction with the next generation of telescopes and their impact on characterizing the chemistry of exoplanetary atmosphere, we will continue to use chemistry to learn about the underlying details of planet formation.

\begin{acknowledgements}

We thank the two anonymous referees who's advice was greatly appreciated, and improved the quality of the manuscript. Thank you to Paul Molli\`ere and Arthur Bosman for their spirited discussions and use of Paul's chemical equilibrium code. Astrochemistry in Leiden is supported by the European Union A-ERC grant 291141 CHEMPLAN, by the Netherlands Research School for Astronomy (NOVA), and by a Royal Netherlands Academy of Arts and Sciences (KNAW) professor prize. The work made use of the Shared Hierarchical Academic Research Computating Network (SHARCNET: www.sharcnet.ca) and Compute/Calcul Canada. R.E.P. is supported by an NSERC Discovery Grant. M.A. acknowledges funding from NSERC through the PGS-D Alexander Graham Bell scholarship.

\end{acknowledgements}


\appendix

\section{ Protoplanetary disk model }\label{sec:disk}

The important backbone to both planet formation and astrochemical studies is the protoplanetary disk model which describes the distribution of both the gas and dust. For the gas we use a modified version of the analytic disk model of \cite{Cham09}, while for the dust we use the Two-population model of \cite{B12}. These models are both outlined below.

\subsection{ Gas model }\label{sec:gas}

The disk model of \cite{Cham09} is a self-similar solution to the diffusion equation assuming that gas accretes towards the host star at a constant rate of \citep{Alb05}:\begin{align}
\dot{M} = 3\pi\nu\Sigma_g,
\label{eq:01}
\end{align}
within a disk radius $r_{\rm switch}$ and away from the host star outside $r_{\rm switch}$ at the same rate to conserve angular momentum. $\Sigma_g$ is the gas surface density, and the viscosity ($\nu$) is given by the standard $\alpha$-prescription \citep{SS73}. With this prescription, the viscosity is $\nu = \alpha c_s H$, where $c_s$ is the gas sound speed, and $H$ is the gas scale height. The source and size of the disk-$\alpha$, which parametrizes the angular momentum transport by either gas turbulence or magnetic winds, is not well constrained and is an ongoing topic of discussion \citep{BaiStone2013,Bai2016,Bai2017,Tazzari2017,Najita2018,Pascucci2018,Wang2018,Milliner2019}. 

As was done in past work (see \citealt{Crid17} \& \citealt{AP18}), we assume that the disk-$\alpha$ is a combination of both turbulence and disk winds: $\alpha = \alpha_{turb} + \alpha_{wind}$ such that it is a constant throughout the whole disk, with $\alpha = 0.001$. In this way, we assume by construction that the gas turbulence and magnetic winds are working in tandem to maintain a constant accretion through the disk. Even when (for example) the gas is passing through a dead zone, where $\alpha_{\rm turb}$ drops (see below).

The gas is heated by two sources, the loss of gravitational energy as material accretes and the direct irradiation by the host star. For the former process the disk effective temperature is given by \citep{CieslaCuzzi2006}:\begin{align}
2\sigma T^4_{\rm eff} = \frac{9\nu\Sigma_g\Omega}{4} = \frac{3\dot{M}\Omega}{4\pi},
\label{eq:02}
\end{align}
where $\sigma$ is the Boltzmann constant and $\Omega = \sqrt{GM_*/r^3}$ is the Kepler frequency. The corresponding midplane temperature is:\begin{align}
T^4 = \frac{3\tau }{4}T^4_{\rm eff},
\label{eq:03}
\end{align}
where the optical depth of the gas is:\begin{align}
\tau = \frac{\kappa_0 \Sigma_g}{2},
\label{eq:04}
\end{align}
and $\kappa_0$ is the gas opacity which we assume to be constant for gas temperatures below the dust sublimation temperature ($\sim 1300$ K). We follow \cite{Cham09} in setting $\kappa_0 = 3 {\rm cm^2~g^{-1}}$.

We scale the mass accretion rate from eq. \ref{eq:01} by the absolute magnitude of the mass accretion rate at the disk edge: \begin{align}
\dot{M}_0 = 3\pi\nu_0\Sigma_0,
\end{align}
and combine eqs. \ref{eq:01}, \ref{eq:02}, and \ref{eq:03} such that the midplane temperature in the viscously accreting region of the disk is:\begin{align}
\left(\frac{T}{T_{vis}}\right)^4 = \left(\frac{\nu}{\nu_0}\right)\left(\frac{\Sigma_g}{\Sigma_0}\right)^2\left(\frac{\Omega}{\Omega_0}\right)^2,
\end{align}
with: \begin{align}
T_{vis} = \left(\frac{27\kappa_0}{64\sigma}\right)\nu_0\Sigma_0^2\Omega_0^2.
\end{align}

Further from the host star (r $> 10$ AU) the gas density is too low for viscous accretion to efficiently heat the disk. Far from the host star the disk is additionally more flared and more easily intersects incoming stellar radiation. Here the absorption and re-radiation of stellar light is the primary source of heating. 

As in \cite{Cham09} we follow the model of \cite{CG97} in the irradiated region of the disk (and assmume T$_{\rm gas}$ = T$_{\rm dust}$) so that the effective temperature of the gas is: \begin{align}
T_{eff} = \left[\frac{r}{2}\frac{d}{dr}\left(\frac{H_{ph}}{r}\right)\right]^{1/4}\left(\frac{R_*}{r}\right)^{1/2} T_*,
\label{eq:08}
\end{align}
where $H_{ph}$ is the height of the photosphere, and $R_*$ and $T_*$ are the radius and effective temperature of the host star. We assme that the stellar properties remain constant throughout the Class-II phase of the proto-stellar system. Assuming that the disk is roughly vertically isothermal then \cite{CG97} found that:\begin{align}
\frac{H_{ph}}{r} \simeq 4\left(\frac{T_*}{T_c}\right)^{4/7}\left(\frac{r}{R_*}\right)^{2/7},
\label{eq:09}
\end{align}
with:\begin{align}
T_c = \frac{GM_*\mu m_H}{kR_*}.
\label{eq:10}
\end{align}
Similarly scaling by the gas properties at the disk outer edge and combining eqs. \ref{eq:08}, \ref{eq:09}, and \ref{eq:10} we find that the midplane temperature is:\begin{align}
\left(\frac{T}{T_{rad}}\right) = \left(\frac{r}{s_0}\right)^{-3/7},
\label{eq:11}
\end{align}
where $s_0$ is the radius of the outer edge of the disk, and: \begin{align}
T_{rad} = \left(\frac{4}{7}\right)^{1/4}\left(\frac{T_*}{T_c}\right)^{1/7}\left(\frac{R_*}{s_0}\right)^{3/7}T_*.
\label{eq:12}
\end{align}

Both regions of the disk occur concurrently, within a certain radius ($r_t$, known as the heat transition) the heating is dominated by viscous accretion while outward of the heat transition the heating is dominated by direct irradiation. Such that the disk midplane temperature is:\begin{align}
T(r) = \begin{cases}
T_{vis}\left(\frac{\nu}{\nu_0}\right)^{1/4}\left(\frac{\Sigma_g}{\Sigma_0}\right)^{1/2}\left(\frac{\Omega}{\Omega_0}\right)^{1/2} & r < r_t, \\
T_{rad}\left(\frac{r}{s_0}\right)^{-3/7} & r > r_t.
\end{cases}
\label{eq:13}
\end{align}

To derive the gas surface density we combine the scaled version of eq. \ref{eq:01} and eq. \ref{eq:13}: \begin{align}
\Sigma_g(r) = \begin{cases}
\Sigma_{vis}\left(\frac{\dot{M}}{\dot{M}_0}\right)^{3/5}\left(\frac{r}{s_0}\right)^{-3/5} & r < r_t, \\
\Sigma_{rad}\left(\frac{\dot{M}}{\dot{M}_0}\right)\left(\frac{r}{s_0}\right)^{-15/14} & r > r_t,
\end{cases}
\label{eq:14}
\end{align}
where $\Sigma_{vis} = \Sigma_0(T_0/T_{vis})^{4/5}$ and $\Sigma_{rad} = \Sigma_0 (T_0/T_{rad})$. The particular value of $\Sigma_0$ and $T_0$ depend on the dominate heating source. If viscous accretion dominates at the outer edge of the disk (i.e. $T_{vis} > T_{rad}$) then $T_0 = T_{vis}$, otherwise $T_0 = T_{rad}$. The disk in the former case is characterized by a single power law and the hypothetical location of the heat transition is beyond the edge of the disk (i.e. $r_t > s_0$).

In the case that $r_t > s_0$ the normalization $\Sigma_{vis} = \Sigma_0$ is determined through the initial disk mass:\begin{align}
M_{disk,0} = \int_0^{s_0} 2\pi\Sigma_g(r,t=0) dr,
\label{eq:15}
\end{align}
while in the case where $r_t<s_0$:\begin{align}
M_{disk,0} &= \int_0^{r_t} 2\pi \Sigma_g(r < r_t,t=0) dr \nonumber\\
& \qquad + \int_{r_t}^{s_0} 2\pi\Sigma_{g}(r > r_t,t=0) dr.
\label{eq:16}
\end{align}
For details see \cite{Cham09}, we find:\begin{align}
\Sigma_0 = \Sigma_{vis} &= \frac{7M_{disk,0}}{10\pi s_0^2} \nonumber\\
\Sigma_{rad} &= \Sigma_0\frac{T_0}{T_{rad}} \nonumber
\end{align}
when $r_t > s_0$ and: \begin{align}
\Sigma_0 = \Sigma_{rad} &= \frac{13M_{disk,0}}{28\pi s_0^2}\left[1- \frac{33}{98}\left(\frac{T_{vis}}{T_{rad}}\right)^{52/33}\right]^{-1} \nonumber\\
\Sigma_{vis} &= \Sigma_0\left(\frac{T_0}{T_{vis}}\right)^{4/5} \nonumber
\end{align}
when $r_t < s_0$.

As the disk evolves and the majority of the mass accretes towards the host star, and this requires the transport of angular momentum away from the inner disk. When viscous accretion dominates the angular momentum transport, the disk outer edge of the disk ($s$) spreads outward from its initial size ($s_0$). Meanwhile if disk winds dominate the evolution, the angular momentum is lifted above the disk as matter flows along magnetic field lines. In this case the outer edge of the disk would not evolve, or would tend to shrink. In our work we assume that the outer disk is dominated by viscous angular momentum transport, and hence the outer edge should grow. The angular momentum of the disk is:\begin{align}
L = 2\pi\sqrt{GM_*}\int_0^s r^{3/2} \Sigma_g(r) dr.
\label{eq:17}
\end{align}

Combining eqs. \ref{eq:15}, \ref{eq:16}, and \ref{eq:17} one can derive the temporal evolution of the mass accretion rate of the disk. They have the form (see \cite{Cham09} for details):\begin{align}
\dot{M}(t) = \begin{cases}
\frac{\dot{M}_0}{\left(1 + t/\tau_{vis}\right)^{19/16}} & t < t_1,\\
\frac{\dot{M}_1}{\left(1 + (t-t_1)/\tau_{rad}\right)^{20/13}} & t > t_1,
\end{cases}
\label{eq:18}
\end{align}
where $t_1$ is the time first time when $T_{rad} > T_{vis}$, and $\dot{M}_1$ is the mass accretion rate at $t_1$. Furthermore $\tau_{vis} = 3M_0/16\dot{M}_0$ and $\tau_{rad} = 7M_1/13\dot{M}_1$ are evolution timescales, and $M_1$ is the mass of the disk at $t_1$. In the case when the outer disk is heated by direct irradiation (i.e. when $T_{rad} > T_{vis}$) at the start of the simulation, then $t_1 = 0$.

Along with its viscous accretion, a gas disk is also susceptible to photoevaporation by Far- and Extreme- UV photons. The process of photoevaporation is complicated and a full treatment (i.e. in \cite{Gort09,Gort15}) is beyond the scope of this work. However we account for photoevaporation using the simplier method outlined in \cite{HP13}. In their model, photoevaporation reduces the global mass accretion rate, by lifting the gas out of the disk, at an exponential rate:\begin{align}
\dot{M}^\prime(t) = \dot{M}(t)\exp{\left(-\frac{t-t_{int}}{t_{dep}}\right)},
\label{eq:19}
\end{align}
for $t>t_{int}$, where $t_{int} = 10^{5}$ yr is the time when the photoevaporation begins to take effect, and $t_{dep}$ is the depletion timescale from photoevaporation. As in \cite{AP18} we set the depletion timescale to be equal to the disk lifetime. In this way disks with higher lifetimes have intrinsically lower photoevaporation rates (by construction) than disks with lower lifetimes.

\subsection{ Dust model }\label{sec:dust}

The distribution and evolution of the dust disk impacts the production and growth of planetesimals via the streaming instability \citep{Schafer17}. Additionally, since the volatile freeze out rate is slower (by mass of dust) for larger grains and larger grains have lower optical depth (per mass of dust) than smaller grains, the average dust grain size also impacts the chemistry in the disk \citep{Jonkheid2006,Jonkheid2007,Vasyunin2011,Fogel11,Crid16a,Krijt2016,Krijt2018}).

The treatment of the dust evolution in our model is computed with the Two-population model of \cite{B12}. This model is a semi-analytic approximation that constructs the grain size distribution and dust surface density to replicate the results of numerical simulations of grain growth and fragmentation. A particularly useful feature of this approach is that it is computationally cheaper to run than full dust simulations, allowing it to be quickly run in support of astrochemical codes and planet formation calculations.

The evolution of the dust is primarily dictated by its Stokes number ($St = a\rho_s\pi / 2\Sigma_g$) which describes how much aerodynamic drag a grain of a certain size ($a$) and density ($\rho_s$) will feel. For a given grain chemical abundance (and hence density) the Stokes number is limited by three processes: its growth timescale, fragmentation, and radial drift. In the fragmentation limited regime, grains will grow to a point where any subsequent collision leads to erosion rather than grow. The analytic calculations of \cite{B11} have shown that the maximum grain size in this regime is:\begin{align}
a_{frag} = f_f\frac{2}{3\pi}\frac{\Sigma_g}{\rho_s\alpha_{turb}}\frac{u_f^2}{c_s^2},
\label{eq:20}
\end{align}
where $c_s$ is the gas sound speed. The parameter $f_f$ is a factor of order unity, meant to take into account the fact that the actual maximum grain size from numerical simulations is {\it slightly} below the maximum size predicted theoretically. A key parameter to this process is the `fragmentation threshold speed' $u_f$. When two grains impact with speeds above $u_f$ they tend to fragment rather than stick together.

Following our past work \citep{Crid16b}, we assume that a dust grain with an ice mantle has a fragmentation threshold speed of 10 m/s. For a dust grain without an ice mantle we assume a fragmentation threshold speed of 1 m/s. This implies a radial dependence on the fragmentation threshold speed that changes across the water ice line - where water changes its state between vapour and ice.

A second method of limiting the size of the dust grain is through the tension between dust growth and radial drift. When a grain's growth timescale:\begin{align}
t_{grow} = a/\dot{a},
\label{eq:21}
\end{align}
exceeds the timescale of radial drift \citep{B12}:\begin{align}
t_{drift} = \frac{r^2\Omega}{Stc^2_s}\gamma^{-1},
\label{eq:22}
\end{align}
then it will drift inward faster than grow larger. This has an expected maximum grain size of \citep{B12}:\begin{align}
a_{drift} = f_d\frac{2\Sigma_d}{\pi\rho_s}\frac{(r\Omega)^2}{c_s^2}\gamma^{-1},
\label{eq:23}
\end{align}
where $\Sigma_d$ is the surface density of the dust and $\gamma$ is the absolute value of the gas pressure gradient. The parameter $f_d=0.55$ is set to properly fit the results of the semi-analytic model to numerical simulations.

Because of the transition in fragmentation threshold speed at the water ice line, the combination of fragmentation and radial drift leads to a build up of dust just inside the water ice line. This build up is caused by a reduction in the average grain size inward of the water ice line. These smaller grains radially drift slower than the grains outward of the ice line - causing a traffic jam \citep{Pinilla2016}.
The increase in dust surface density within the ice line works to increase the efficiency of planetesimal formation and the further growth of planets \citep{Drazkowska2016}.

Another important transition point is at the outer edge of the dead zone. The gas turbulence is generally believed to be generated by the magnetorotational instability (MRI, see for ex. \citealt{BaiStone2013}), which requires a high enough electron fraction ($n_e/n_H\sim 10^{-10}$) for the magnetic field to couple to the bulk gas. At high enough densities, both cosmic rays and high energy photons can not penetrate enough to supply this level of ionization \citep{BS11,Gressel2015}. Hence the MRI saturates and the turbulence level drops by about two orders of magnitude \citep{Crid16b}.

As was done in \cite{Crid16b} we have included the effect of an evolving dead zone by dropping $\alpha_{\rm turb}$ to $10^{-5}$ within the dead zone. There, the maximum grain size typically becomes dominated by radial drift, since turbulent driven fragmentation becomes much less efficient. We define the location of the dead zone edge through a parametrized model which was developed in \cite{Crid16b}.  The edge location depends on the ionization structure of the disk gas, as computed by our astrochemical model. Our model assumes a maximum of singly charged dust grains which could underestimate the amount of charge held by the grains, depending on the average size of the dust grains.

\subsection{ UV and X-ray photon field }\label{sec:field}

Along with its impact on planet formation, the dust distribution strongly impacts the flux of high energy radiation. These high energy photons ionize the gas and drive fast chemical reactions in the gas phase. As was done in \cite{Crid16b} we combine our dust model described in section \ref{sec:dust} with the radiative transfer scheme {\it RADMC3D} \cite{RADMC}, following its internal dust opacity calculator to determine the impact of the changing dust distribution on the radiation field. The optical constants use in the opacity calculations were taken from \cite{Draine03}.

For use in our chemical calculation (described below) we pre-compute both the dust radial and size distribution of the dust grains along with the resulting radiation field for a number of different times through the disk. Each of these snapshots are separated in time by on average $10^5$ years. 

Our general method to compute the physical properties of the gas (temperature, density), dust (size, density), and radiation field is to first generate a snapshot of the gas properties at every time step from a starting time of $t_0 = 10^5$ yr to the end time $t_{final} = t_{life}$. Next we use the two-population model described above to compute the size and radial distribution of the dust, and assume that it has the same temperature as the gas. Finally we use the gas and dust distributions as inputs to {\it RAMDC3D} to compute the high energy radiation field.

With these spatial distributions in hand we compute the chemical structure of the disk for each snapshot (see section \ref{sec:chem}).

\section{ Planet formation }\label{sec:plntform}

\subsection{ Planetary growth: Planetesimal accretion }\label{sec:acc}

The population used in this work is a subset of a population of synthetic planets from Alessi, Pudritz, \& Cridland (APC, in prep). Our subset includes their Zone 1 planets (also known as hot-Jupiters/Neptunes), defined by final orbital radii $r < 0.1$ AU and planetary mass $M_p \geq 10$ M$_\oplus$; and close-in Zone 5 planets (classically called super-Earths or sub-Neptunes) with $r<0.1$ AU and $M_p < 10$ M$_\oplus$. 

The background model used in APC is featured in \cite{AP18}, and was built using the planetesimal accretion paradigm which posits that planets grow through the sequential collisions of large (10-100 km) bodies followed by the accretion of gas after the proto-planet has reached a mass of $\sim 10$ M$_\oplus$. These planetesimals would hypothetically be constructed through the efficient coagulation of pebble-sized grains driven by the streaming instability (see for ex. \citealt{Schafer17}).  We do not, however account for any reduction in the surface density of dust due to the build up of planetesimals. Such an addition would be useful to be consistent with observations of dust mass in protoplanetary disks \citep{Tychoniec2018}, however we leave it to future work.

This scenario is based on the prescription developed by \cite{KI02} and \cite{IL04a}, and begins with a seed mass of 0.01 M$_\oplus$ that accretes planetesimals on a timescale of \citep{IL04a}:\begin{align}
t_{\rm acc} &= 1.2\times 10^5 {\rm yr} \left(\frac{\Sigma_d}{10 {\rm g~cm}^{-2}}\right)^{-1}\left(\frac{r}{1~{\rm AU}}\right)^{1/2}\left(\frac{M_c}{M_\oplus}\right)^{1/3}\nonumber\\
&\times \left(\frac{M_*}{M_\odot}\right)^{-1/6}\left[\left(\frac{\Sigma_g}{2.4\times 10^3 {\rm g~cm}^{-2}}\right)^{-1/5} \left(\frac{r}{1~{\rm AU}}\right)^{1/20}\left(\frac{m}{10^{18}{\rm g}}\right)^{1/15}\right]^2,
\label{eq:24}
\end{align}
where $M_c$ and $M_*$ are the mass of the proto-planet and host star respectively, while $m$ is the assumed (constant) mass of each of the incoming planetesimals. With this timescale, the proto-planetary mass grows as: $\dot{M}_c = M_c / t_{acc}$.

At a certain point the proto-planet has accreted or scattered all of its surrounding planetesimals. With the reduction of its associated accretion heating the core becomes sufficiently cool to begin capturing gas. We denote this `isolation mass' 
\citep{AP18}: \begin{align}
M_{iso} = 10M_\oplus f_{crit}\left(\frac{\dot{M}_c}{10^{-6}M_\oplus/{\rm yr}}\right)^{1/4},
\label{eq:25}
\end{align}
where the parameter $f_{crit}$ depends on the opacity of gas surrounding the proto-planet \citep{Ikoma2000}: \begin{align}
f_{crit} = \left(\frac{\kappa_{env}}{1 {\rm cm^2g^{-1}}}\right)^{3/10}.
\label{eq:26}
\end{align}

\cite{AP18} explore different values of $\kappa_{env}$ and its effect on the resulting population. However for this work we choose $\kappa_{env} = 0.001 {\rm ~cm^2 g^{-1}}$ as it represented the population of planets that best reproduced the population of known exoplanets.

As gas accretes it must shed its gravitational energy to remain bound to the planet. The incoming gas cools on the Kelvin-Helmholtz timescale \citep{Ikoma2000}:\begin{align}
t_{KH} = 10^c {\rm yr}\left(\frac{M_p}{M_\oplus}\right)^{-d},
\label{eq:27}
\end{align}
where $M_p$ is the mass of the growing planet and the parameters $c$ and $d$ depend on $\kappa_{env}$. Using the expressions from \cite{AP18} (their eqs. 23 and 24) and $\kappa_{env} = 0.001 ~{\rm cm^2 g^{-1}}$: $c=7.7$ and $d=2$. Assuming that the Kelvin-Helmholtz timescale is the limiting factor to gas accretion then the gas accretion rate is: \begin{align}
\dot{M}_p = M_p/t_{KH}.
\label{eq:28}
\end{align} 

While the planet initially accretes gas slowly, as its mass builds it eventually begins a stage of unstable gas accretion. As a final varied parameter in the population synthesis model we set a maximum mass above which a planet will stop its growth. As in \cite{AP18} the maximum mass is:\begin{align}
M_{\rm max} = f_{\rm max}M_{\rm gap},
\label{eq:29}
\end{align}
where the parameter $f_{max}$ is stochastically varied between 1-500 and $M_{gap}$ is the gap opening mass. As defined in \cite{MP06} and \cite{AP18}:\begin{align}
M_{\rm gap} = M_*{\rm min}\left(3h^3,\sqrt{40\alpha_{\rm turb}h^5}\right),
\label{eq:30}
\end{align}
where $h=H/r$ and $H = c_s/\Omega$ is the gas scale height. The first term in eq. \ref{eq:30} denotes the point when the planet's hill radius exceeds the scale height of the disk. At this point the planet's gravitational effect exceeds the dynamics of the gas.  The second term denotes the mass of the planet when its gravitational torque exceeds the viscous torque of the surrounding gas. We assume that when either of these requirements are met that a gap in the disk is opened.

The premise of setting a randomly selected upper limit to the total mass of the planet is a required simplification because of our lack of knowledge of the mechanism responsible for terminating growth. This `terminal mass problem' was discussed by \cite{Morbidelli2014} who connected the end of planetary growth to gap opening, however found that gas flow into the gap could still be maintained through meridional circulation from heights above the midplane. Recently \cite{Batygin2018} hypothesised that this circulating flow could be prevented from accreting onto the planet through complex interactions with the planet's dynamo-driven magnetic field. With that in mind, \cite{Cridland2018} derived a semi-analytic function for the gas accretion rate, and indeed found that accretion could be stopped with this magnetic shielding. \cite{Cridland2018} derived limits of $f_{\rm max}$ and found a range between $\sim$ 100-400 for a limited sample of forming planets. Hence our choice in the range of $f_{max}$ used in this population is reasonable when compared to this physically motivated model.

Since the population used here is pre-computed by APC we stick to the termination prescribed in eq. \ref{eq:29} and leave the effect on the population of magnetic termination to future work.

\subsection{ Planetary migration: Planet traps }\label{sec:trap}

Before it opens a gap, an embedded planet perturbs the surrounding gas as it grows. Numerical simulations (see for ex. \citealt{Masset2002}) have shown that these perturbations lead to spiral waves in the disk gas that back-react on the planet, predominantly removing its angular momentum. Known as the Lindblad torque, these spiral waves cause the planet to migrate inward, towards the host star. The Lindblad torque is balanced by the co-rotation torque, that is caused by material orbiting on horseshoe orbits around the planet \citep{Masset06,Paard10,Lyra2010,Crid16a}.

For a simple disk model like the minimum mass solar nebular (MMSN, \citealt{Crida2009}) the combined effect of the Lindblad and co-rotation torque is inward migration. However more complicated disk models can include transitions in disk properties that strengthen the co-rotation torque producing locations of net zero torque - these are known as `planet traps'. Outlined in \cite{HP11}, three such planet traps exist in typical protoplanetary disks. They are the water ice line, heat transition, and the edge of the dead zone. These traps represent transitions in the dust opacity by the freeze-out of water, a change in the heating mechanism (discussed above), and a change in the turbulent strength respectively. More recently, \cite{Crid19a} used models of astrochemistry, dust opacity, and radiative transfer to rigorously show that these traps do indeed lead to a zero point in the net torque on a planet. As in \cite{AP18} we will assume that a growing planet is completely trapped in one of these planet traps until they open a gap in the disk.

Once the planet mass exceeds the gap opening mass (eq. \ref{eq:30}) it opens a gap in the gas disk. The planet now begins migrating via Type II migration since the torques responsible for its original Type I migration are switched off by the presence of the gap. Type II migration proceeds at the viscous timescale ($t_\nu = r^2/\nu$), with the planet acting as an intermediary to the global angular momentum transport of the disk. This process continues until the mass interior to the planet falls below the current planetary mass. At which point Type II migration stalls (see \cite{APC16a} for details).

\bibliographystyle{aa} 
\bibliography{mybib.bib} 

\end{document}